\documentclass[12pt]{article}
\usepackage{amsmath}
\usepackage{amssymb}
\usepackage{amsthm}
\usepackage{amsfonts}
\usepackage{color}
\usepackage{graphicx}
\usepackage{booktabs}
\usepackage{graphicx,epstopdf}
\usepackage{mathtools}
\usepackage{mathrsfs}
\usepackage{multirow}
\usepackage{rotating}
\usepackage{enumerate}
\usepackage{epsfig}
\usepackage{setspace}
\usepackage{ragged2e}
\usepackage{blindtext}
\usepackage{lineno}
\usepackage{url}
\usepackage{caption}
\usepackage{subcaption}
\usepackage{physics}
\usepackage{lscape}
\usepackage{float}
\usepackage{authblk}
\usepackage[colorlinks=true,linkcolor=red,citecolor=blue]{hyperref}%

 \usepackage[style=authoryear, backend=biber, maxcitenames=1, maxbibnames=99]{biblatex}
\theoremstyle{definition}

\usepackage{geometry}

\numberwithin{equation}{section}

\title 
 {\bfseries Towards improved pest management of the  soybean aphid}
 \author[1]{Urvashi Verma}
 \author[2]{Margaret Lewis}
 \author[2]{Jordan Lehman}
 \author[1]{Rana D. Parshad}
\affil[1]{Department of Mathematics, Iowa State University, Ames, IA 50011, USA.}
 \affil[2]{Department of Entomology, Ohio State University, Columbus, OH 43210, USA.}

\begin{document}
\date{}
\maketitle


\begin{abstract}
The soybean aphid (\emph{Aphis glycines}) is an invasive insect pest that continues to cause large-scale damage to soybean crops in the North Central United States. The current manuscript proposes several mathematical models for the top-down bio-control of the aphid, as well as control via pesticides and neonicotinoids. The models are motivated empirically, and constructed based on laboratory experiments conducted to test control of aphids by Lacewing larvae, as well as by a parasitic wasp (\emph{Aphidius colemani}). The effectiveness of these models is compared by taking into account factors such as economic injury levels for soybeans, life history traits such as cannibalism amongst the predator, and intraguild predation between competing bio-control agents such as predators and parasitoids. The models predict multiple population peaks and transient chaotic dynamics when a predator and/or insecticides are used. It is observed that parasitoids, in conjunction with predators, are more efficient at stabilizing the population dynamics than insecticide use. They also suggest a combination of predators, parasitoids, and insecticides would be more efficient at suppressing aphid populations than using only predators or parasitoids. The models also qualitatively capture the features seen in long-time field data from 2000-2013. We discuss applications of our results to pest management strategies for soybean aphids in the context of a changing climate, as well as regime shifts.

\end{abstract}

\vspace{1em}

\section{Introduction}

The invasive, pestiferous soybean aphid (\emph{Aphis glycines}) was first detected in the United States in Wisconsin in July 2000 (\cite{ragsdale2004soybean}). 
Since that time, it has become a significant pest of soybeans in the North Central United States, causing yield losses as high as 40\% when unchecked (\cite{song2006profitability,kim2008economic}) and triggering a significant increase in chemical insecticide usage within soybean production (\cite{ragsdale2011ecology}). 
 In addition to raising production costs (\cite{song2006profitability}), this increased reliance on insecticides raises concerns about environmental risks (\cite{bahlai2010choosing}), such as agrochemical run-off as well as negative impacts to beneficial insect populations, including biological control agents (\cite{desneux2004effects, desneux2007sublethal,johnson2008preventative}). Indeed, soybean aphids face a diverse array of endemic natural enemies that includes predators (\cite{rutledge2004soybean}), parasitoids (\cite{kaiser2007hymenopteran}), and entomopathogenic fungi (\cite{nielsen2005control}). These biological control agents have the capacity to significantly suppress soybean aphid populations (\cite{fox2004predators,costamagna2006predators,clifton2018effects,kaser2018impact}), although this does not occur consistently, and predicting the efficacy of biological control services remains challenging. To begin addressing this critical knowledge gap, a deeper understanding of the population dynamics between soybean aphids and their natural enemies is needed. Quantifying these interactions, as well as the biotic and abiotic factors that modulate them, may be a critical step for increasing reliance on biological control within soybean integrated pest management programs. Performing experiments that explore these interactions over a long period of time, such as an entire growing season, can be logistically challenging, limiting our ability to consider how pest and natural enemy dynamics might shift under varying scenarios. Mathematical modeling provides an important tool for bridging this gap.

\vspace{0.2cm}

\par Mathematical models have previously been used to describe \emph{Aphis glycines} population dynamics. Analyzing complex interactions between aphids and their natural enemies, these models provide useful information about population dynamics and the effectiveness of a biological control strategy. Several studies have used such models to explore the aphid population dynamics (\cite{kindlmann2004simple,costamagna2007exponential, matis2007stochastic,matis2009population}) and their regulation by natural enemies (\cite{kindlmann2010modelling}). Houdkov{\'a} et al. (\cite{houdkova2006scaling}) present a model to understand the long-term dynamics of aphids and an aphidophagous predator. Their findings indicate that the top-down regulation fails in ladybird–aphid systems, on a metapopulation scale - the impact of predators on aphid average density (across many years) is relatively small. Matrix population models are also used to model stage-structured populations. 
Both (\cite{lin2003effect}) and (\cite{miksanek2019matrix}) applied stage-structured matrix models to analyze the dynamics of aphid-parasitoid interactions. However, the parasitoid species studied in (\cite{lin2003effect}) was 
\emph{Aphidius colemani} while (\cite{miksanek2019matrix}) focused on \emph{Aphelinus certus}. Mills in (\cite{mills2005selecting}) also use a stage-structured matrix model to evaluate the importance of parasitism at different stages in the life cycle of the codling moth. 

\vspace{0.2cm}

\par Nicholson and Bailey were among the first to model host–parasitoid interactions using a discrete-time model for non-overlapping generations (\cite{nicholson1935balance}). A large body of literature has since built upon this framework (\cite{mills1996modelling,hassell2000spatial,jang2012discrete,murdoch2013consumer,livadiotis2015discrete}), though there are still gaps that need to be addressed. In systems with overlapping host generations, such as soybean aphids, a continuous time model framework is more suitable (\cite{ives1992continuous}). 
Delay differential equations also provide a valuable framework to model host-parasitoid interactions, accounting for time lags due to development or maturation (\cite{murdoch1987invulnerable,briggs1999delayed,wearing2004stage}), but, this modeling approach has not been incorporated in this manuscript. Existing host-parasitoid models do not explicitly capture the boom-bust dynamics exhibited by soybean aphids in nature. While some models have explored aphid dynamics with predators and parasitoids separately, few have modeled both types of bio-control agents acting in conjunction (\cite{snyder2003interactions,nakazawa2006community,huang2022mathematical}). To the best of our knowledge, models exhibiting boom-bust dynamics along with both predators and parasitoids, as well as insecticide application, in a continuous time framework, have not been considered in the literature. However, capturing these interactions is crucial for improving biocontrol strategies aimed at suppressing aphid outbreaks.  To address the limitations in aphid biocontrol models to date, this manuscript presents several continuous-time population models that capture complex interactions between soybean aphids and their natural enemies. 

To increase the ecological relevance of our simulations, we performed controlled laboratory experiments to generate various model parameters, including age-stage survival rates in aphids, predator functional responses, stage specific parasitoid attack rates, and parasitoid emergence rates from mummified aphids. We used these models to explore various complex aspects of aphid - natural enemy interactions, including long-term (\textit{i.e. multi year}) dynamics, comparisons of different natural enemy population dynamics, the impact of insecticides on these interactions, and the extent to which different model settings limit aphid populations from reaching either the economic threshold (ET) or economic injury level (EIL) within soybeans (\cite{ragsdale2007economic} ). Although future work will be needed to validate and explore these models, particularly within the context of multi-species interactions, this work collectively highlights the potential for biological control to limit soybean aphid populations and allows us to identify scenarios where biological control might be most effective.

\section{Materials and Methods}
 
\subsection{Experimental Evaluation of Predator and Parasitoid Population Dynamics}

To support model development, we conducted laboratory and greenhouse assays that quantified various aspects of the interactions between the soybean aphid and its natural enemies. Experiments were conducted using two natural enemies: green lacewing larvae (\textit{Chrysoperla rufilabris}) and a parasitoid wasp (\textit{Aphidius colemani}). Both species are known biocontrol agents for the soybean aphid, and they can be purchased from commercial beneficial insectaries. All experiments were conducted using an avirulent (biotype 1) soybean aphid colony that was continuously reared on soybean for over three years, an \textit{Aphidius colemani} colony that had been maintained on soybean aphids for approximately one year, and commercially purchased green lacewing larvae that were reared from eggs to the third instar on soybean aphids (see supplemental materials for further details on the insects used in these experiments). The results from these experiments were used to guide model development as well as the selection of model parameters used in the simulations below.

\subsubsection{Quantifying Predation Rates in Lacewing Larvae}
\label{Quantifying Predation Rates in Lacewing Larvae}
We conducted greenhouse mesocosm assays to quantify lacewing predation rates on soybean aphid nymphs. Soybean plants in the V3 growth stage (i.e., three completely unfolded trifoliates) were infested with $50$ fourth instar soybean aphid nymphs, with nymphs distributed evenly between the trifoliates. Each plant was then placed into an individual $47.5$ cm$^3$  BugDorm cage, and aphids were allowed to settle for approximately $60$ minutes before introducing lacewing larvae. Plants were infested with either one $3^\text{rd}$ instar lacewing larvae or three  $3^\text{rd}$ instar lacewing larvae ($\sim  7$ days old). For plants that only received one lacewing, the larvae were placed on the top trifoliate. For plants that received three lacewings, larvae were distributed across all three trifoliates. After introducing the lacewings, plants and insects were held in a temperature-controlled growth chamber for four hours, at which point we froze the plants and counted the number of aphid nymphs and lacewing larvae remaining. Lacewing consumption rates were compared between plants with one and three lacewing larvae using a general linear model in the R Base package with a binomial distribution. Models were checked for overdispersion by comparing the Pearson chi square to the residual degrees of freedom, and models were not overdispersed. Six replicates were performed.

We observed no differences in aphid consumption rates due to lacewing density $(\text{X}^2 = 1.85, \text{DF} = 1,10, \text{P} = 0.173; \text{Table} \ \ref{table:predator_experiment})$. In mesocosms with three lacewing larvae, we observed some evidence of cannibalism among the predators. Three of the plants that were originally inoculated with three lacewing larvae were missing one or two lacewing larvae at the end of the four-hour experimental period.

\begin{singlespace}
\begin{table}[H]
\centering
\begin{tabular}{|l|c|c|c|}
 \hline
\footnotesize{\textbf{Initial Lacewing Density} }& \footnotesize{\textbf{Mean $\#$ Aphids Consumed}} $\pm$ \footnotesize{\textbf{SE (\%)}} & \footnotesize{\textbf{Mean $\#$ Lacewings after 4h.}} & \footnotesize{\textbf{N}} \\
\hline
1& 25.67 $\pm$ 3.15 & 1.00 $\pm$ 0.00 & 6\\
\hline
3 & 30.67 $\pm$ 4.63 & 2.20  $\pm$ 0.37 & 6\\
\hline
\end{tabular}
\caption{The mean and standard error for the number of aphids consumed and the number of lacewing larvae remaining on plants after the four-hour experimental period.}
\label{table:predator_experiment}
\end{table}
\end{singlespace}
\subsubsection{Parasitism Impacts on Aphid Fecundity}
\label{Parasitism Impacts to Aphid Fecundity}

We conducted laboratory assays to evaluate how parasitoid attacks from \textit{Aphidius colemani} impact soybean aphid fecundity. Avirulent aphid adults were either exposed to one mated female \textit{A. colemani} until a confirmed parasitoid attack occurred (“parasitized”) or left unexposed as a control (additional experimental details provided in the Supplemental Methods 
\ref{supp_materials}). 
Aphids were then moved onto individual petri dishes with a single leaf and monitored daily for survival and nymph production. “Parasitized” aphids were monitored until death or mummy formation occurred (approximately $6$ to $7$ days). “Control” aphids were monitored for $10$ days, the maximum amount of time required for mummy formation based on previous development assays (Lewis and Lehman, Personal Communication). $12$ experimental replicates were performed in total. We summarized data as the cumulative daily nymph production, and differences between treatments were analyzed using repeated measures ANOVA. Time (days since the start of the experiment), parasitoid exposure treatment, and their interaction were included as fixed effects. Subject ID (\textit{i.e.} denoting individual aphids) nested within a day was also included as a random effect to account for repeated measures on each experimental unit.

\vspace{0.2cm}

\begin{figure}
\centering
\includegraphics[width=7cm, height=5cm]{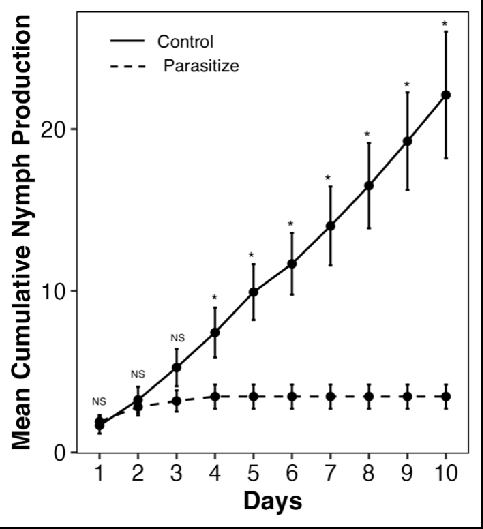}
    \caption{Mean cumulative nymph production  \( \pm \) standard error in parasitized and unparasitized soybean aphids. Asterisks denote days that had significant differences in cumulative nymph production between parasitized and unparasitized soybean aphid nymphs (alpha = 0.05).}
    \label{fig:exp_para_1}
\end{figure}

Soybean aphid fecundity changed in response to the interaction between parasitoid exposure and time $(\text{F}_{1,20.9} = 38.8, \text{P} < 0.001)$. Aphids continued to produce offspring for up to four days after a parasitoid attack, with no significant differences in fecundity compared to control aphids noted during days $1-3$. Starting on the day $4^{\text{th}}$, we began to observe a significant decrease in cumulative nymph production within the parasitized aphids (Figure \ref{fig:exp_para_1}).
\subsubsection{Evaluating Parasitoid Preference for Different Aphid Instars}
\label{Evaluating Parasitoid Preference for Different Aphid Instars}
We conducted petri dish assays to determine if \textit{Aphidius colemani} host preference varied between different soybean aphid instars. Briefly, three soybean leaves were placed into a large ( $ \sim  10$ cm height $ \times 15$ cm diameter) petri dish (Supplemental Figure 
\ref{fig:supp_2}). 
Leaves were then infested with 10 soybean aphids belonging to one of three juvenile groups: A) Juvenile Group $1$ (first/second instars), B) Juvenile Group $2$ (third/fourth instars), or C) Adults. Aphids were pulled directly from a laboratory colony, and we distinguished between juvenile groups based on relative body size, the size/color of their cornicles, and the presence of nearby offspring (to confirm the adult life stage). To minimize bias in aphid selection, all aphids were selected by the same person. Aphids were allowed to settle on the leaves for $\sim 10$ minutes, at which point we introduced one female parasitoid wasp from a laboratory colony into the petri dish. Using a $0.95$ mL gel capsule, we placed the female wasp into the center of the assay arena so that she was equidistant from all three leaves. Parasitoid wasps were not age-standardized, but we did monitor the dishes to ensure that the female attacked an aphid within the first ten minutes of the assay. Once an attack successfully occurred, we left the wasp and aphids undisturbed for two hours. The parasitoid was removed after two hours. Aphids from each leaf were transferred onto caged plants and held for $10$ days, at which point we counted the number of mummies that formed. For the second and third blocks, we also placed any mummies that formed into individual gel capsules and monitored for adult emergence. Data were summarized as the attack rate (\textit{i.e.} proportion of aphids in each instar group that became mummies) and the adult emergence rate (\textit{i.e.} proportion of mummies that successfully produced an adult parasitoid wasp). In total, $10$ experimental replicates were performed over three separate experimental blocks. Data were analyzed using a binomial GLM model in the R v4.4.2 base package  \href{https://www.r-project.org/}{(R Core Team 2024)}, with the response variable consisting of the ratio of parasitized aphids to unparasitized aphids across each treatment. The Aphid age group was included as an explanatory variable, and we checked the ratio for the residual degree of freedom to Pearson’s Chi-Square to confirm that the model was not overdispersed. 
\begin{singlespace}
\begin{table}[H]
\centering
\begin{tabular}{|l|c|c|}
\hline
\small{\textbf{Aphid Age Group}} & \small{\textbf{Mean Attack Rate}} $\pm$ \small{\textbf{SE (\%)}} & \small{\textbf{Mean Adult Emergence Rate}} $\pm$ \small{\textbf{SE (\%)}} \\
\hline
First / Second & 0 $\pm$ 0 & N/A \\
\hline
Third / Fourth & 27.67 $\pm$ 11.17 & 58.75 $\pm$ 15.60 \\
\hline
Adults & 46.71 $\pm$ 13.14 & 56.67 $\pm$ 13.22 \\
\hline
\end{tabular}
\caption{Variation in parasitoid attack rate (proportion of aphids that successfully form mummies) and adult emergence (proportion of aphid mummies that successfully produce a live parasitoid adult wasp) as a function of aphid age.}
\label{aphid_rates}
\end{table}
\end{singlespace}

Parasitoid attack rates varied across the different juvenile group treatments, with the highest rates of mummy formation occurring in larger soybean aphid hosts $(\text{X}^{2} = 70.48, \text{DF} = 2, 27, \text{P} < 0.001; \text{Table} 
 \ \ref{aphid_rates});$. No parasitoid mummies formed in the first /second instar nymphs across any of our experimental replicates, while the highest rates of mummy formation occurred in the adult aphids. 
Among the aphid age groups that successfully produced mummies, no differences in adult parasitoid emergence rates were noted $(\text{X}^{2} = 0.09, \text{DF} = 2, 7, \text{P} =0.76; \text{Table} \ \ref{aphid_rates})$.

\subsection{Mathematical models for pest management of Soybean Aphids}
Drawing from the experimental results of sections \eqref{Quantifying Predation Rates in Lacewing Larvae}, \eqref{Parasitism Impacts to Aphid Fecundity} and,  \eqref{Evaluating Parasitoid Preference for Different Aphid Instars} in the next few subsections, various mathematical models will be developed and proposed to incorporate the effect of predators, parasitoids, insecticides and combinations thereof, on soybean aphids. All the simulations presented in the results were performed using the solver \href{https://www.mathworks.com/help/matlab/ref/ode45.html}{ode$45$} in \href{https://www.mathworks.com/help/matlab/ref/ode45.html}{MATLAB}.

\subsubsection{Single-species models for Aphid population dynamics}
(\cite{kindlmann2010modelling}) introduced a model to describe the population dynamics of aphids using a set of differential equations,
\begin{equation}
\begin{split}
\label{eq:0}
\frac{dh}{dt} &= a x, \ h(0) =0 \\
\frac{dx}{dt} &=  (r-h)x, \ x(0) =x_{0}\\
\end{split}
\end{equation}

where $h(t)$ refers to the cumulative population density of aphids and $x(t)$ refers to the aphid population density at time $t$. $a$ is the scaling constant and $r$ is the growth rate of aphids. The aphid population initially increases due to the linear growth term, corresponding to the ``boom'' phase. But as the cumulative density $h$ surpasses the growth rate $r$, the population starts declining because of competition, giving the ``bust'' phase, see Figure \ref{kindlmann_model_img}. The model has been thoroughly analyzed and exhibits an exact solution, (\cite{matis2007stochastic,matis2009population}). Various other single-species models that describe the dynamics of \textit{Aphis glycines} have also been proposed
 (\cite{costamagna2007exponential}). Herein, the growth rate is modeled as a decreasing function of time, which can switch from positive to negative, after a critical time at which the peak of the population occurs. The model has also accurately represented the population dynamics of \textit{A. glycines} across various conducted experiments (\cite{costamagna2007exponential}). However, in order to be consistent, we choose \eqref{eq:0} to model aphid dynamics henceforth.
\subsubsection{Predator bio-control with stage structure and  type II response}
 Generally, all stages of predators eat aphids indiscriminately. However, we structure the predator population into juveniles and adults to accurately represent the lacewing predator from the experiments in section \ref{Quantifying Predation Rates in Lacewing Larvae}, which only preys on aphids when it is in its juvenile stage. The adult life stage is not predaceous, with adults instead feeding on nectar, honeydew, and pollen (\cite{limburg2001extrafloral}). Beyond lacewings, this split in adult and larval feeding behavior is seen in other generalist predators, including hover flies \emph{(Syrphidae)} and midges \emph{(Itonididae)}(\cite{brodeur2017predators}). Thus, it is important to incorporate stage structure in the predator population from a modeling perspective to accurately represent the biology.

\vspace{0.2cm}

\par In our proposed model \eqref{eq:03}, the assumption about the aphid population is that its dynamics follows \eqref{eq:0}, via the classical cumulative density model (\cite{kindlmann2010modelling}) and is thus governed by \eqref{eq:0}. Here, $x(t)$ refers to the aphid population density, and $h(t)$ is the cumulative population density of aphids at time $t$. $y_j(t)$ refers to the predator population density in the juvenile stage, and $y_a(t)$ represents the predator population density in the adult stage. A general description of the model parameters is given in Table \ref{table_pred_params}. 

\begin{singlespace}
\begin{table}[H]
    \centering
    \begin{tabular}{|l l|}
        \hline
        \textbf{Parameters} & \textbf{Definition} \\
        \hline
        \multicolumn{2}{|l|}{} \\
        $a$    &  scaling parameter relating aphid cumulative density to its dynamics\\
        $r$    & maximum potential growth rate of aphids\\
        $\xi$  & birth rate of new juveniles from adult predators\\
        $\beta_j$ & rate at which juvenile predator matures to the adult stage\\
         $\epsilon_x$ & conversion efficiency for aphids \\
        $a_x$ & attack rate of juvenile predators on aphids \\
        $t_x$ & handling time spent for aphids \\
        $\delta_a$  & natural mortality of adult predators \\
        $c$ &   decay rate constant of insecticide \\
$\delta_k$ & initial efficacy of insecticide on group $k$\\
        \hline
    \end{tabular}
   \caption{Parameter definitions for model \eqref{eq:03} and \eqref{eq:insectide04}} 
    \label{table_pred_params}
\end{table}
\end{singlespace}
\label{pred_section_1}
\begin{equation}
\begin{split}
\label{eq:03}
\frac{dh}{dt} &= a x, \ h(0) =0 \\
\frac{dx}{dt} &=  (r-h)x -\textcolor{black}{\dfrac{ a_xxy_{j}}{1+(a_xt_x)x}}, \ x(0) =x_{0}\\
\frac{dy_{j}}{dt} &= \textcolor{black}{\dfrac{\epsilon_x a_xxy_{j}} {1+(a_xt_x)x}} + \textcolor{black}{\xi y_a} -\textcolor{black}{\beta_j y_j}, \ y_{j}(0) =y_{j_{0}}\\
\frac{dy_{a}}{dt} &=  \textcolor{black}{\beta_j y_j} - \delta_a y_{a}, \ y_{a}(0) = y_{a_{0}}\\
\end{split}
\end{equation}
\par The aphid population is depredated on only by the juvenile stage of the predator (lacewing larvae). Thus, the predation term appearing in the aphid population equation depends only on $y_j$, and there is no involvement of adult predators $y_a$. This feeding supports the growth of lacewing larvae $y_j$ with the conversion efficiency $\epsilon_x$, which can be seen in the juvenile predator population equation. This feedback to the predator via $\epsilon_x$ separates our model from several biocontrol models in the literature (\cite{houdkova2006scaling}), where it is stated that the predators do not reproduce on the time scale (such as a summer season) for aphid control. However, with lacewings, it is well known that there will be several generations (even within one summer season) (\cite{gerling1997dynamics,umn_lacewing}), and so one needs to accurately represent this feedback in a biocontrol model, with lacewings. Also, the growth of $y_j$ depends not only on aphid consumption but also on the reproduction by adult predators $y_a$. The juvenile predator eventually matures to the adult stage, contributing to the growth of the adult predator population $y_a$. 

\vspace{0.2cm}

\begin{figure}[H]
\centering
\includegraphics[width = 7cm,height=5cm]{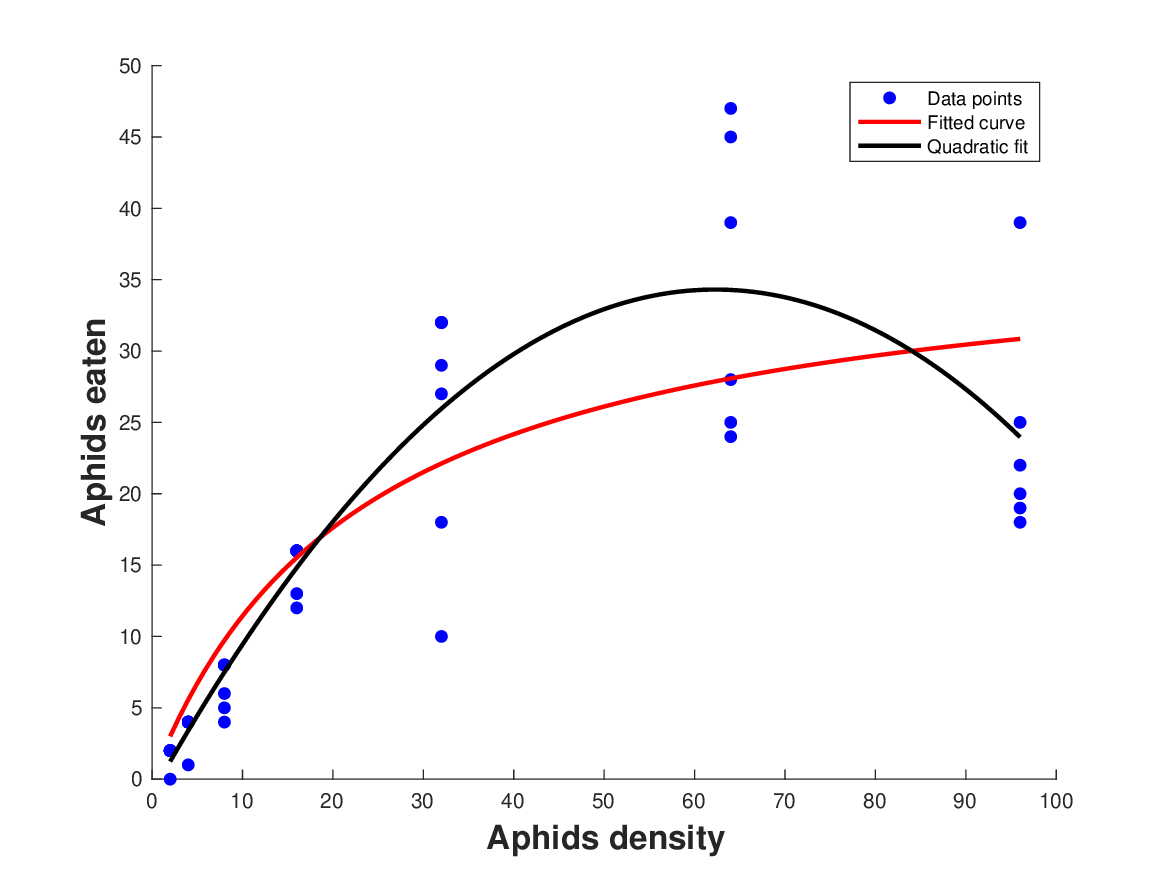 }
\caption{The blue dots represent the data points observed in laboratory experiments. The red curve corresponds to a fitted model using the `nlinfit' function in MATLAB, and the black curve represents a quadratic fit obtained by using the `polyfit' function with a polynomial degree of 2 in MATLAB software.}
\label{fig:curve_fit}
\end{figure}

Preliminary data from the experiments conducted to determine functional response suggest that the predators exhibit a type II response, which can be seen in Figure \ref{fig:curve_fit} through the red curve. The Matlab function `nlinfit' is used for nonlinear regression fitting to estimate two crucial parameters from the data: (i) the attack rate of juvenile predators on aphids $a_x$ and (ii) the handling time spent for prey (aphids) $t_x$. Using this function and experimental data, we obtain $ a_x=1.6281 \ \text{and} \ t_x=0.0260$, which will be kept the same throughout this manuscript for predator attacking on aphids. We also compare the curve fitting errors obtained from `nlinfit' and `poly fit'. The three main error metrics calculated are Root Mean Square Error (RMSE), Mean Absolute Error (MAE), and the degree of linear correlation ($\text{R}^{2}$). Table \ref{table:curve_fitting_errors} shows the results for comparison. Even though the quadratic polynomial fit shows better error metrics, it lacks biological meaning and has overfitting risk, while the type II functional response with nlinfit provides a more reliable description of predator-prey interactions. Since cannibalism is also evident with lacewing larvae, the experiment was conducted to evaluate how predation upon conspecifics impacts their consumption rates on aphids and conspecifics (Supplemental section 
\ref{Impact of Cannibalism from Conspecific Predators}). 
Previously, mathematical models considered cannibalism in predators (\cite{houdkova2006scaling}), but here, the model accounts for the positive growth in predator population from feeding on aphids, see model \eqref{eq:04}
in Supplemental Methods. 
\begin{singlespace}
\begin{table}[H]
 \centering
\begin{tabular}{|l|c|c|}
\hline
\textbf{Error Metric} & \textbf{Nonlinear regression (nlinfit)} & \textbf{Polynomial curve fitting (quadratic)} \\
\hline
RMSE  & 6.6389  & 5.3153 \\
 \hline
MAE   & 4.7353  & 3.5079 \\
 \hline
$R^2$ & 0.7252  & 0.8238 \\
\hline
\end{tabular}
\caption{Comparison of error metrics for Holling type II using nlinfit and quadratic polynomial curve fitting.}
\label{table:curve_fitting_errors}
\end{table}
\end{singlespace}

\subsubsection{Predator bio-control with insecticide treatment}
\label{pred_with_insecticide}

Insecticides have been extensively used to control the soybean aphid population since their introduction (\cite{bahlai2015shifts, ragsdale2011ecology}). To model the use of insecticide, one way is to introduce specific functions that behave like foliar application of insecticides, that result in aphid mortality. These can be sprayed multiple times in a growing season. $I_k$ represents the cumulative insecticide effect on a population group $k$ at time $t$. The negative effect of insecticides is incorporated via functions $I_1$ for aphids, $I_2$ for juvenile predators, and  $I_3$ for adult predators.  Herein, the decay rate $c$ was obtained from the half-life of various insecticides in the literature.
Applications are modeled via the use of an indicator function $\mathbf{1}_{\{ t \geq t_{\text{spray},i} \}}$, which includes only past sprays in the summation. The intensity of insecticides' impact $(\delta_k)$  can vary between aphids and predators. Thus, each insecticide spray has an immediate effect that gradually wears off over time.
Also, via this framework, we are able to explore one application per season, such as a neonicotinoid, or several applications per season, such as traditional sprays/foliar applications, or even combinations thereof. General description of the model parameters is given in Table \ref{table_pred_params}.
\begin{equation}
\begin{split}
\label{eq:insectide04}
\frac{dh}{dt} &= a x, \ h(0) =0 \\
\frac{dx}{dt} &=  (r-h)x - I_1x - \textcolor{black}{\dfrac{ a_xxy_{j}}{1+(a_xt_x)x}}, \ x(0) =x_{0}\\
\frac{dy_{j}}{dt} &= \textcolor{black}{\dfrac{\epsilon_x a_xxy_{j}} {1+(a_xt_x)x}} + \textcolor{black}{\xi y_a} - I_2{y_j}-\textcolor{black}{\beta_j y_j}, \ y_{j}(0) =y_{j_{0}}\\
\frac{dy_{a}}{dt} &=  \textcolor{black}{\beta_j y_j} - \delta_a y_{a}-I_3{y_a}, \ y_{a}(0) = y_{a_{0}}\\
\text{Where } I_k &= \delta_k \sum_{i: t \geq t_{\text{spray},i}}\!e^{-c(t - t_{\text{spray},i})} \mathbf{1}_{\{ t \geq t_{\text{spray},i} \}}, k =1,2,3\\
\end{split}
\end{equation}

\subsubsection{Parasitoid bio-control with stage structure in the aphid population}
For parasitoids, the aphid acts as a host for their larvae development. During this process, parasitoid larvae develop inside the aphid and kill the aphid host upon emergence. 
Since parasitoids may prefer specific life stages of the host for attack (\cite{lin2003effect}), a stage-structured model of aphids is formulated under parasitoid attack. The experimental data drives the model, providing crucial parameters such as the transfer rates from one stage to the next, the attack rate of parasitoids on different life stages, and the emergence rate of parasitoids from mummified aphids for each stage.
\begin{singlespace}
\begin{table}[H]
    \centering
    \begin{tabular}{|l l|}
 \hline
 \textbf{Parameters} & \textbf{Definition} \\
\hline
\multicolumn{2}{|l|}{} \\
 $a$    &  scaling parameter relating aphid cumulative density to its dynamics\\
 $r$    & maximum potential growth rate of aphids\\
  $\alpha$  & coefficient for the impact of parasitoids on aphid fecundity\\
 $\beta_1, \beta_2$ & maturation rate of aphids from stage $j_1$ to $j_2$ and from stage $j_2$ to the adult stage $a$\\
$\eta_1, \eta_2$ & parasitoid emergence rates from mummified aphids in stages $j_2$ and $a$,  respectively\\
$\gamma_1, \gamma_2$ & attack rate of the parasitoid on the ${j_2}$ stage  and adult stage of aphids, respectively\\
 $\delta$  & natural death rate of parasitoids \\
 \hline
\end{tabular}
\caption{Parameter definitions for model \eqref{eq:05}} \label{table_parasitoid_params}
\end{table}
\end{singlespace}
\begin{equation}
\begin{split}
\label{eq:05}
\frac{dh}{dt} &= a (x_{j_{1}}+x_{j_2}+x_a), \ h(0) =0\\
\frac{dx_{j_1}}{dt} &=  \dfrac{rx_a}{1+\alpha p}- \beta_1x_{j_1},  \ x_{j_{1}}(0) =x_{j_{1_0}}\\
\frac{dx_{j_2}}{dt} &= \beta_1x_{j_1} - \beta_2x_{j_2}-  \dfrac{\gamma_1 x_{j_2}p}{1+x_{j_2}}, \ x_{j_{2}}(0) =x_{j_{2_0}}\\
\frac{dx_{a}}{dt} &=  \beta_2x_{j_2}-h(x_{j_{1}}+x_{j_2}+x_a)-  \ \dfrac{\gamma_2 x_{a}p}{1+x_{a}},  \ x_{a}(0) =x_{a_{0}}\\
\frac{dp}{dt} &= \dfrac{\eta_{1} x_{j_2} p}{1+x_{j_2}} + \dfrac{\eta_{2} x_{a} p}{1+x_{a}} - \delta p, \ p(0) =p_{0}\\
\end{split}
\end{equation}
\par In this model \eqref{eq:05}, the aphid population density in the first juvenile stage $j_1$ involves $1^\text{st}$ and $2^\text{nd}$ instar is represented by $x_{j_1}(t)$  and $x_{j_2}(t)$ denotes the aphid population density in the second juvenile stage $j_2$, including $3^\text{rd}$ and $4^\text{th}$ instar. $x_a(t)$ refers to the aphid population density in the adult stage, and $h(t)$ is the combined cumulative population density of all the stages at time $t$. The parasitoid population density is denoted by $p$.  Through experimental data (section  \eqref{Evaluating Parasitoid Preference for Different Aphid Instars}), it was seen that parasitoids primarily target the adult stage, followed by the $j_2$ stage, and no preference was seen for the $j_1$ stage, so the attack rate for the $j_1$ stage is not included in the model. To account for this stage-specific parasitism, the model includes different attack rates by parasitoid and also allows for different emergence rates from various aphid stages. The adult aphids are responsible for reproducing new aphids, which contributes to the $j_1$ stage. After a parasitoid attack, the aphid becomes mummified, which reduces its ability to reproduce, reflected in the aphid's growth rate. The maturation rate from each stage contributes to the development of the next stage in the aphid's life cycle, such as from $j_1$ to $j_2$ and from $j_2$ to the adult stage. The assumption here is that overcrowding will most affect the adult stage as limited nutrient availability decreases both their survival and fecundity, due to which the negative effect of the cumulative population is showing up in the aphid's adult stage equation. In the model, the functional response of parasitoids is represented by Holling type II  for both stages of aphids. This choice is supported by empirical studies that suggest many parasitoid species, particularly \emph{Aphidius colemani}, exhibit  Holling type II response (\cite{zamani2006temperature,van1995behaviour}).  A general description of the model parameters is given in Table \ref{table_parasitoid_params}. 

\subsubsection{Predator-Parasitoid coexistence bio-control with insecticide treatment}

For model \eqref{eq:09}, we have implemented the coexistence of predator and parasitoid, and them acting in unison on the aphid population. We allow for intraguild predation as well, whereby the predator can consume mummified aphids, indirectly killing the parasitoid. Due to the presence of parasitoids, the stage structure in the aphid population is incorporated similar to \eqref{eq:05}. The ${j_1}$ and ${j_2}$ stages are now combined and presented as a single juvenile compartment, because adding up the stages, the maturation rates cancel out. Thus, $x_{j_2}(t)$ now denotes the juvenile aphid population density across all four instars. $x_a(t)$ refers to the aphid population density in the adult stage, and $h(t)$ is the combined cumulative population density of all the stages at time $t$. The parasitoid population density is denoted by $p$. To simplify the model further, the juvenile and adult predator stages are now combined, canceling up the transfer rates. The predator will now grow upon feeding on aphids, and $y_j$ represents the total predator population in the model. Two new stages $x_{m_2}(t)$ and $x_{m_a}(t)$ in the aphid population are introduced, where $x_{m_2}(t)$ denotes the mummified aphid in $j_2$ stage and $x_{m_a}(t)$ refers to the mummified aphid in the adult stage. The reason is that once aphids are parasitized, they turn into mummified aphids and will also be susceptible to attack from predators (\cite{colfer2001predation, costamagna2007suppression}). 
\begin{singlespace}
\begin{table}[H]
\centering
\begin{tabular}{|l l|}
\hline
\textbf{Parameters} & \textbf{Definition} \\
\hline
\multicolumn{2}{|l|}{} \\
$a$  &  scaling parameter relating aphid cumulative density to its dynamics\\
$r$    & maximum potential growth rate of aphids\\
$\alpha$  & coefficient for the impact of parasitoids on aphid fecundity\\
$\beta_1$ & maturation rate of aphids  from stage $j_2$ to the adult stage $a$\\
$\beta_2, \beta_3$ & rates at which parasitized aphids move from $j_2$ to $m_2$ and from $a$ to $m_a$, respectively\\
$\eta_1, \eta_2$ & parasitoid emergence rates from mummified aphids in stages $j_2$ and $a$, respectively\\
$\gamma_1, \gamma_2$ & attack rate of the parasitoid on the ${j_2}$ stage  and adult stage of aphids, respectively\\
 $\epsilon_1, \epsilon_2$ & conversion efficiency for juvenile stage aphids and adult stage aphids, respectively\\
$a_1, a_2$ & attack rate of predators on juvenile stage aphids and adult stage aphids, respectively\\
$t_1, t_2$ & handling time spent for juvenile stage aphids and adult stage aphids, respectively\\
$q_1, q_2$ & probability of attacking the juvenile stage aphids and adult stage aphids, respectively\\
$\delta_p,\delta_j$  & natural death rate of parasitoids and predators respectively\\
$c$ &   decay rate constant of insecticide \\
$\delta_k$ & initial efficacy of insecticide on group $k$ \\
\hline
\end{tabular}
\caption{Parameter definitions for model \eqref{eq:09}} \label{table_para_pred_params}
\end{table}
\end{singlespace}
\begin{small}
\begin{equation}
\begin{split}
\label{eq:09}
\frac{dh}{dt} &= a (x_{j_2}+x_a+x_{m_2}+x_{m_a}), \ h(0) =0\\[0.4em] 
\frac{dx_{j_2}}{dt} &= \dfrac{rx_a}{1+\alpha p}  - \dfrac{\gamma_1 x_{j_2}p}{1+x_{j_2}}- {\dfrac{ q_1a_1x_{j_2}y_{j}}{1+(a_1t_1)\left[(1-q_1)x_{m_2}+q_1x_{j_2}\right]}}-\beta_1x_{j_2} - I_1 x_{j_2}, \ x_{j_{2}}(0) =x_{j_{2_0}}\\[0.4em] 
\frac{dx_{m_2}}{dt} &= \beta_2x_{j_2}-  {\dfrac{ (1-q_1)a_{1}x_{m_2}y_{j}}{1+(a_{1}t_{1})\left[(1-q_1)x_{m_2}+q_1x_{j_2}\right]}} - I_1 x_{m_2}, \ x_{m_{2}}(0) =x_{m_{2_0}}\\[0.4em] 
\frac{dx_{a}}{dt} &=  \beta_1x_{j_2}- h(x_{j_{2}}+x_{m_2}+x_{a}+x_{m_a})-\dfrac{\gamma_2 x_{a}p}{1+x_{a}}- {\dfrac{q_2a_2x_{a}y_{j}}{1+(a_2t_2)\left[(1-q_2)x_{m_a}+q_2x_{a}\right]}} - I_1 x_{a}, \ x_{a}(0) =x_{a_{0}}\\[0.4em] 
\frac{dx_{m_a}}{dt} &= \beta_3x_{a}-  {\dfrac{ (1-q_2)a_{2}x_{m_a}y_{j}}{1+(a_{2}t_{2})\left[(1-q_2)x_{m_a}+q_2x_{a}\right]}}- I_1 x_{m_a},  \ x_{m_a}(0) =x_{m_{0}}\\[0.4em] 
\frac{dy_{j}}{dt} &= \dfrac{\epsilon_1 a_1(q_1x_{j_2}+(1-q_1)x_{m_2})y_{j}} {1+(a_1t_1)((1-q_1)x_{m_2}+q_1x_{j_2})} + \dfrac{\epsilon_{2} a_{2}(q_2x_{a}+(1-q_2)x_{m_a}) y_{j}} {1+(a_{2}t_{2})((1-q_2)x_{m_a}+q_2x_{a})} - \delta_j y_j - I_2 y_j , \ y_{j}(0) =y_{j_{0}}\\[0.5em] 
\frac{dp}{dt} &= \dfrac{\eta_{1} x_{m_2} p}{1+x_{m_2}} + \dfrac{\eta_{2} x_{m_a} p}{1+x_{m_a}} - \delta_p p - I_3 p, \ p(0) =p_{0}\\
\text{Where } I_k &= \delta_k  \sum_{i: t \geq t_{\text{spray},i}} e^{-c(t - t_{\text{spray},i})} \mathbf{1}_{\{ t \geq t_{\text{spray},i} \}}, k =1,2,3\\
\end{split}
\end{equation}
\end{small}
\normalsize

The predators will make a choice at any given time if they will be attacking the aphids in the $j_2$ stage or the $m_2$ stage, to model this, we introduce a probability $q_1$ that predators will attack the $j_2$ stage and then consequently $(1-q_1)$ for the attack on $m_2$ stage. The same framework is applied to the adult and mummified adult stages. This behavior also modifies the type II functional response for the predator, as now it depends on both $x_{j_2}$ and $x_{m_2}$ aphid densities with their respective probabilities.  When predators attack aphids in the ${m_2}$ stage, that interaction affects not only the aphid but also the parasitoid larvae that were growing inside the aphid, so in an indirect way, the predator also affects the parasitoid population. The aphid in the juvenile stage will mature to the adult stage, but the parasitized aphid will move to the mummified compartment $x_{m_2}$, so there will be different transfer rates used to represent those rates. The same logic applies to the aphids in the adult stage $x_a$ as they will move to the mummified adult compartment $x_{m_a}$ after parasitoidal attack. The insecticide effect is modeled in the same way as in model \eqref{eq:insectide04} and explained in section \eqref{pred_with_insecticide}. The assumption is that all four stages of aphids are equally affected by the sprays via $I_1$. However, the predator and parasitoid will also be affected by insecticides with $I_2$ and $I_3$ respectively, but not to the same intensity as aphids. 
A general description of the model parameters is given in Table \ref{table_para_pred_params}. 

\section{Results}
\subsection{Dynamics of model \ref{eq:0}}
\begin{figure}[H]
\begin{subfigure}{.32\textwidth}
\centering
\includegraphics[width=5.8cm, height=5cm]{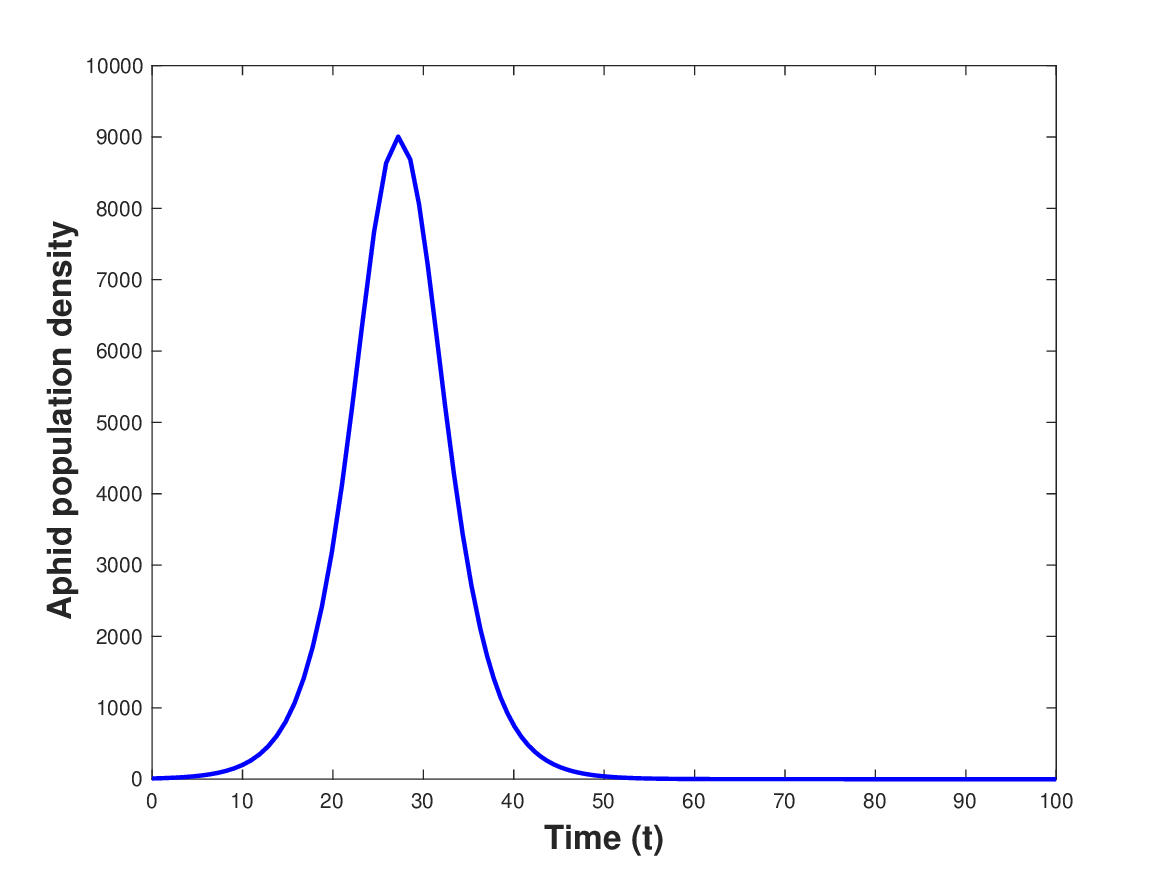}
\subcaption{ $a = 5 \cross 10^{-6}
$ }
\label{kindlmann_model_img}
\end{subfigure}
\begin{subfigure}{.32\textwidth}
\centering
\includegraphics[width=5.8cm, height=5cm]{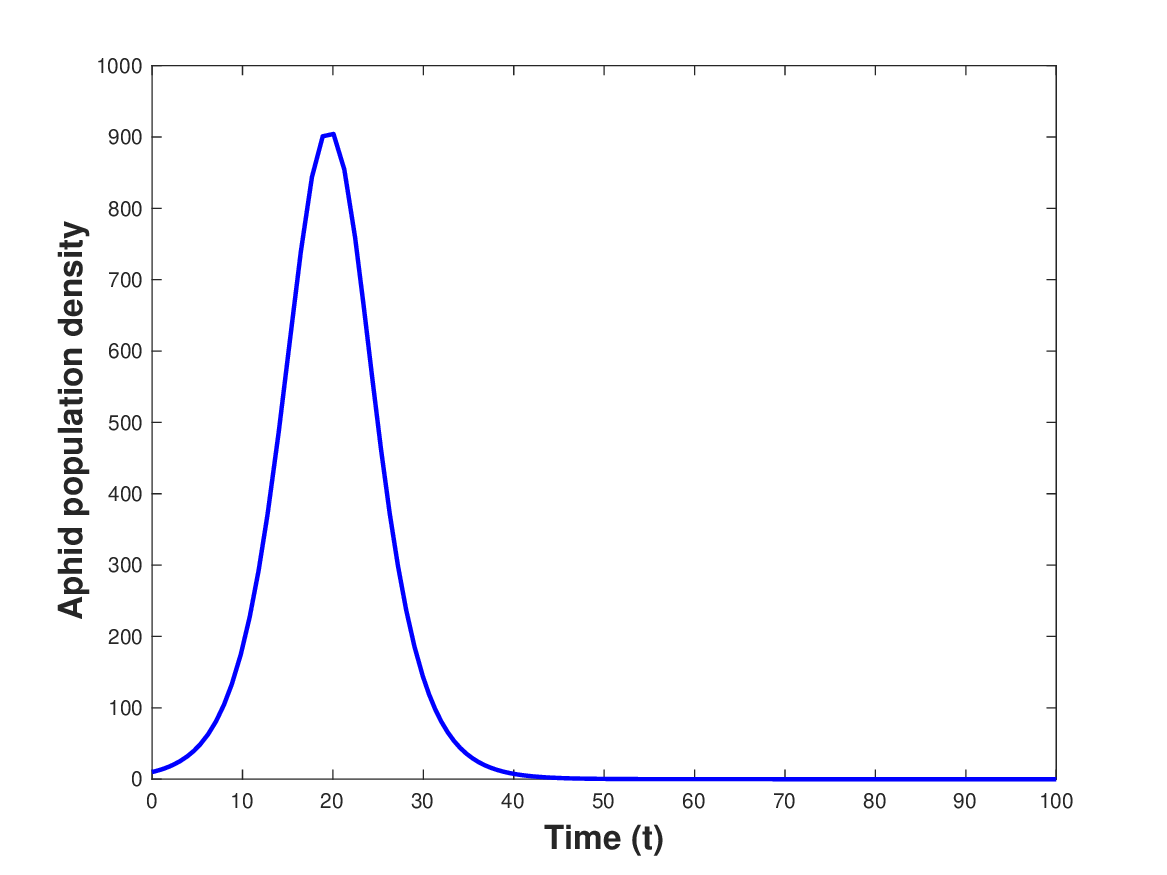}
\subcaption{ $a = 5 \cross 10^{-5}
$ }
\label{kindlmann_model_img_2}
\end{subfigure}
\begin{subfigure}{.32\textwidth}
\centering
\includegraphics[width=5.8cm, height=5cm]{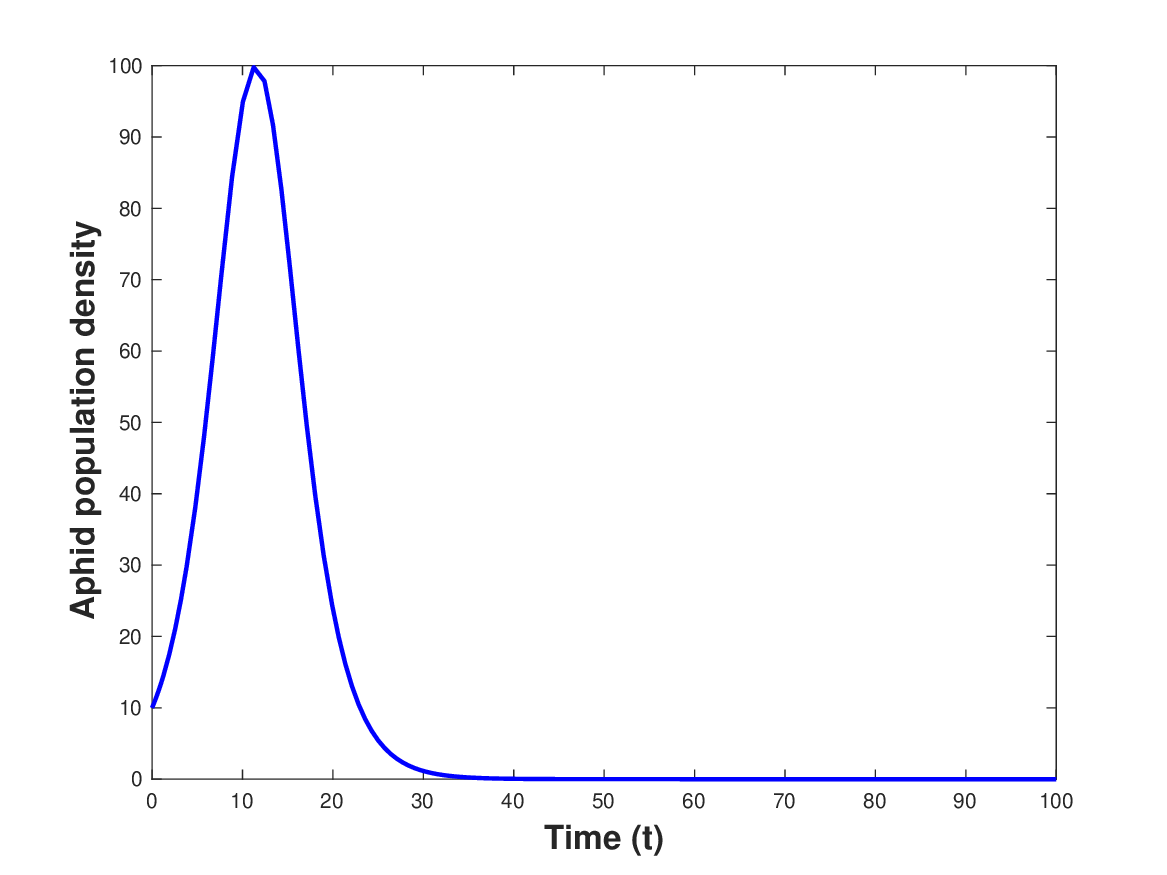}
\subcaption{ $a = 5 \cross 10^{-4}
$ }
\label{kindlmann_model_img_3}
\end{subfigure}
 \caption{The initial population of aphid $x_0$ is set to 10 aphids per plant. The time series subplots show the boom-bust dynamics with respect to varying values of the scaling parameter $a$. With increasing values of $a$, the peak population decreases to a substantial level. Also the unit of time(t) in these simulations, and others hence forth is ``days".}
\label{plot_classical_models}
  \end{figure}

Figure \ref{plot_classical_models} represents the time series for the model \eqref{eq:0}, which corresponds to a single-species aphid model with no control. The growth rate and initial aphid density are kept fixed with $r=0.3, x_{0}=10$ for all the figures \ref{kindlmann_model_img}, \ref{kindlmann_model_img_2}, and \ref{kindlmann_model_img_3}. 
  
\subsection{Results for model \ref{eq:03}}

\begin{figure}[H]
\begin{subfigure}{.3\textwidth}
\centering
\includegraphics[width=5.5cm, height=5cm]{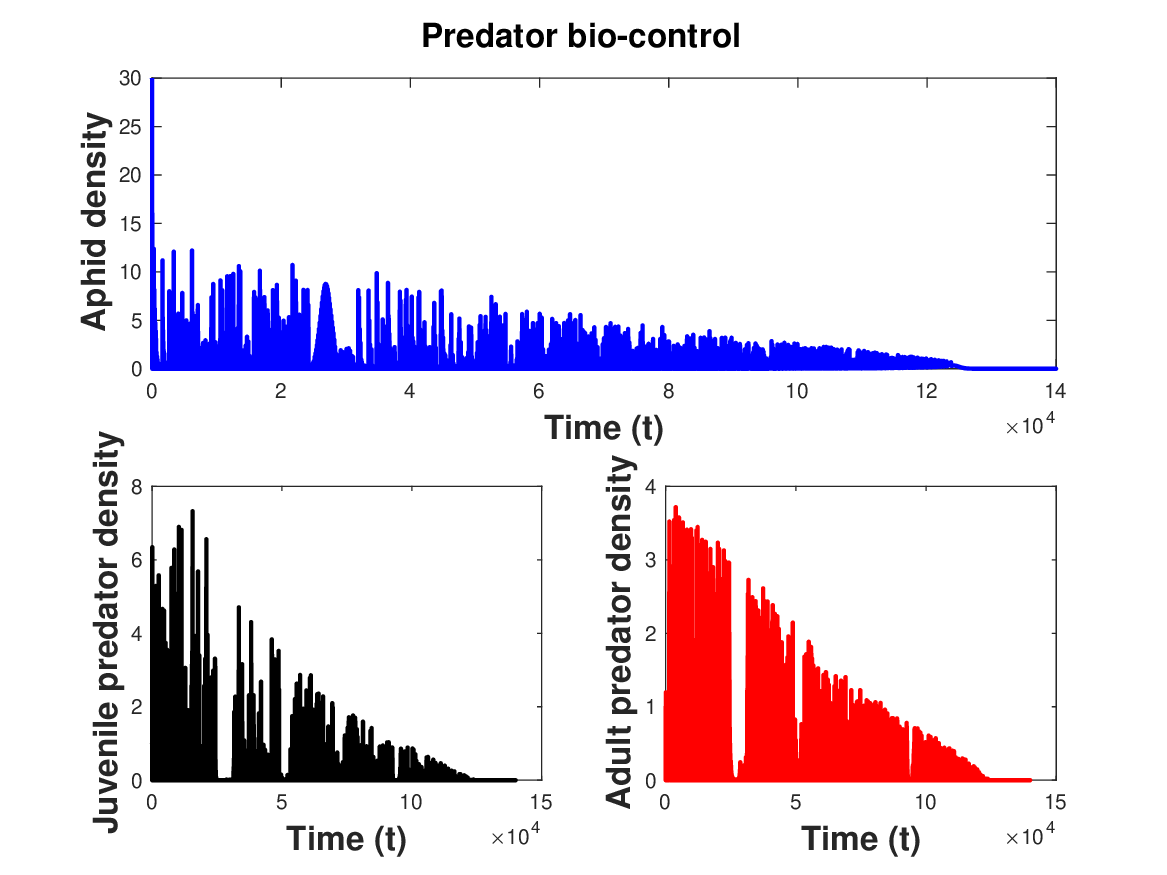}
\subcaption{$a = 5 \cross 10^{-6}
$}
\label{fig:pred7}
  \end{subfigure}
  \begin{subfigure}{.37\textwidth}
  \centering
  \includegraphics[width= 5.7cm, height=5cm]{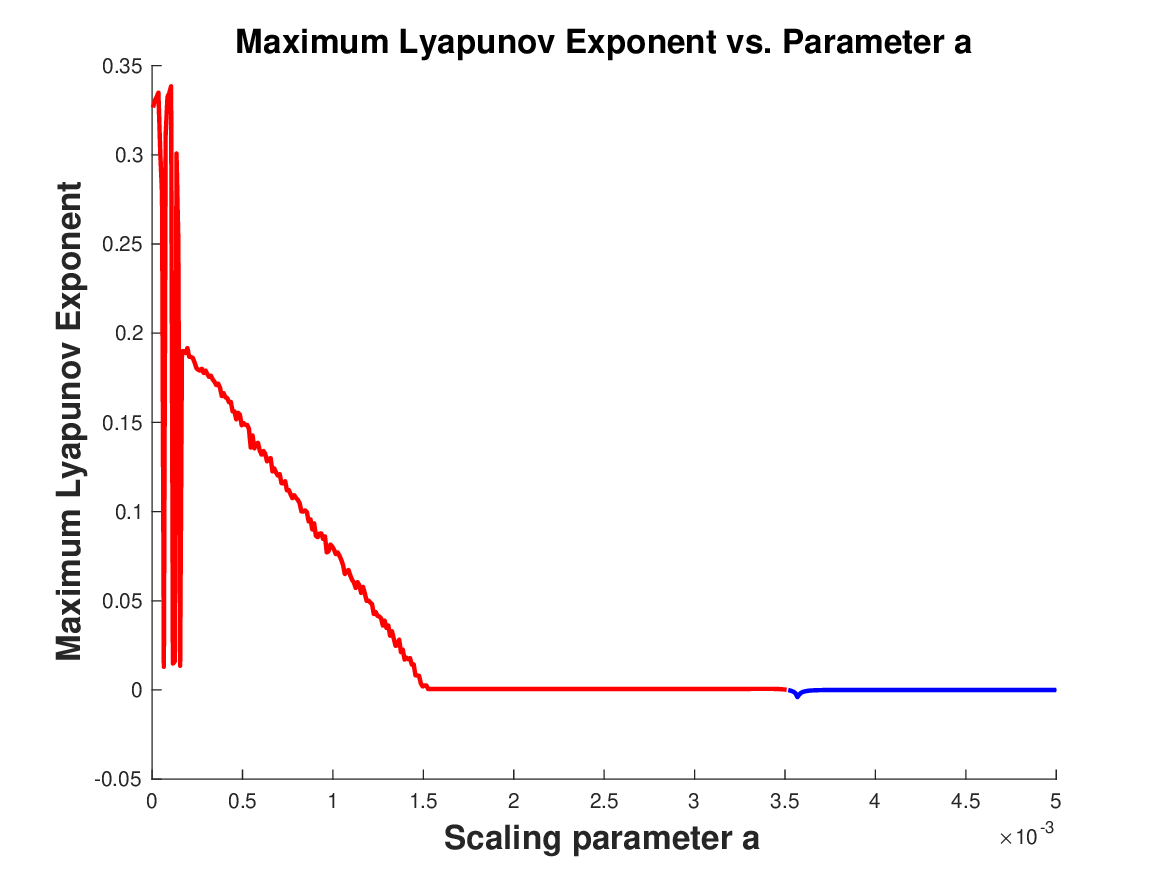}
  \subcaption{$ \text{ Max. Lyapunov exponent across}\ a $}
\label{fig:lyapanov1}
 \end{subfigure}
 \begin{subfigure}{.3\textwidth}
  \centering
  \includegraphics[width= 5.5cm, height=5cm]{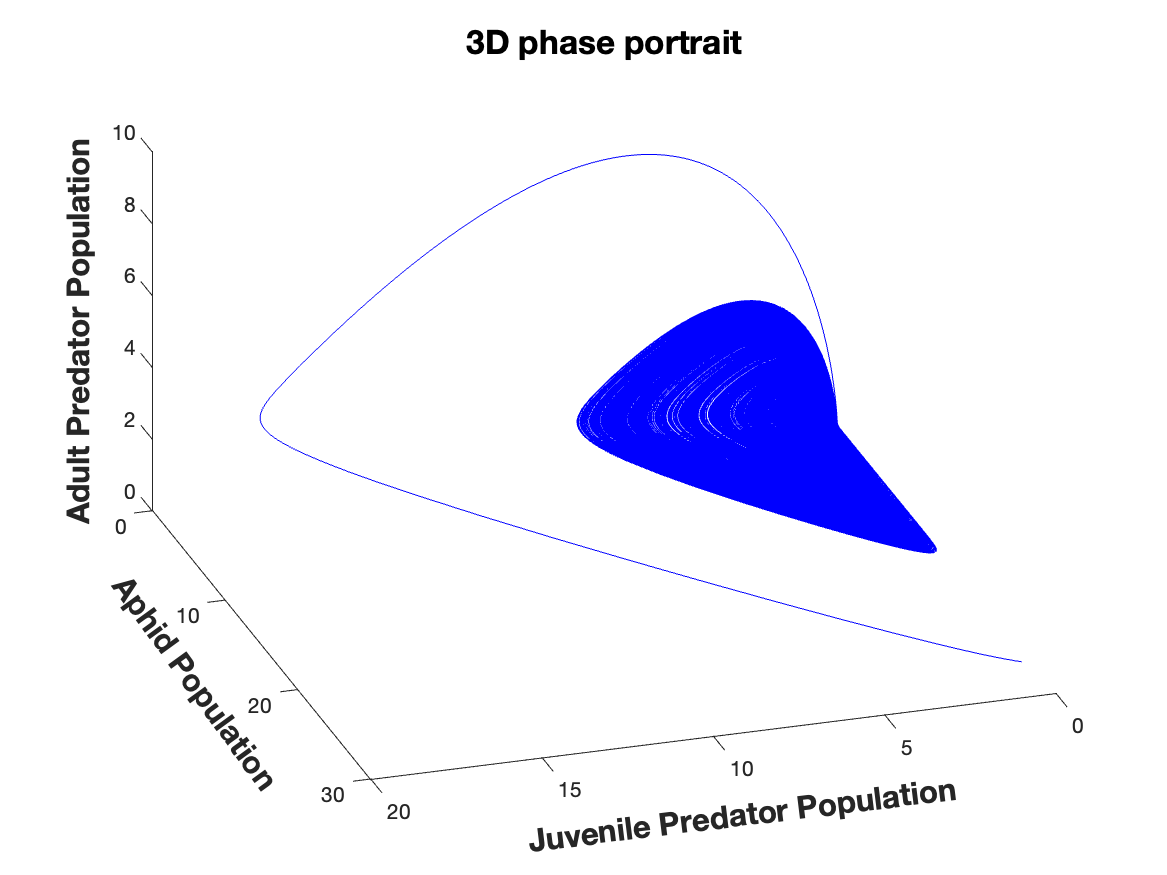}
  \subcaption{$a = 5 \cross 10^{-6}
$}
\label{fig:attractor1}
 \end{subfigure}
 \caption{Figure \ref{fig:pred7} represents long-term time series plots of population densities for model \eqref{eq:03}. The top plot shows the aphid population density over time,  exhibiting chaotic oscillations for a long time period before reaching zero. The bottom left plot displays the juvenile predator population density, which closely follows aphid fluctuations with time-lagged peaks and then declines as the aphid population diminishes. The bottom right plot shows the adult predator population, which also fluctuates due to the maturation of juveniles at a given rate before declining to extinction. Figure \ref{fig:lyapanov1} represents the maximum Lyapunov exponent as a function of the scaling parameter `$a$' for model \eqref{eq:03}. The graph shows chaotic dynamics at some values of `$a$', where the maximum Lyapunov exponent is positive, indicated by red color. The negative Lyapunov exponent, represented by the blue color, corresponds to stable behavior. The phase portrait in Figure \ref{fig:attractor1} shows prolonged chaotic transients, with extinction showing up in the time series Figure \ref{fig:pred7} due to the delayed collapse. The axes represent the aphid population, the juvenile predator, and the adult predator population. The parameters used are $ r=0.3, e_x=0.55, a_x=1.6281,t_x= 0.0260,\beta_j=0.4,\xi=0, \delta_a=0.3  $ with I.C.$=[0,30,1,1]$.}
\label{plot_pred_structure}
 \end{figure}

We observe interesting dynamics when a predator is considered to control aphid populations, via model \ref{eq:03}. The dynamics of the populations seem to behave ``chaotically" - that is the populations fluctuate seemingly randomly, with no order or symmetry to their peaks (see Figure \ref{fig:pred7}). Chaos is defined as aperiodic behavior, exhibiting sensitive dependence to initial conditions (\cite{devaney2018introduction, ott2002chaos}). In our case, the boom-bust dynamics will eventually cause extinction of the aphid populations, so what is observed is actually transient chaos. This is a phenomenon in multi-component ODE systems (typically involving three or
more ODEs), characterized by the population fluctuating chaotically for a finite (but possibly long) period of time, before eventually being attracted to a stable state (\cite{lai2011transient,devaney2018introduction, mccann1994nonlinear}). In the transition phase, similar to chaotic dynamics, aperiodic fluctuations are seen. Also small variations in initial conditions can lead to drastically different outcomes later in time. However, eventually the populations are attracted to the extinction state.

There are various methods to determine transient chaos; we chose to use the standard approach of Lyapunov exponents. In general, when the maximum Lyapunov exponent is positive, this can be considered an indication of chaotic behavior (\cite{ott2002chaos}). Negative and zero Lyapunov exponents indicate that the system exhibits asymptotic stability and neutral stability, respectively. Figure \ref{fig:lyapanov1} represents the maximum Lyapunov exponent with respect to the scaling parameter `$a$'. The parameter `$a$' varies from 5E-6 to 5E-3. It can be seen that the maximum Lyapunov exponent is positive for smaller values of $a$ represented by red color, and turns negative as $a$ increases. When the parameter value `$a=0.000005$' is used, all populations (aphids, juvenile predators, and adult predators) in Figure \ref{fig:pred7} exhibit chaotic transient dynamics over the long time period before eventually dying out. Figure \ref{fig:attractor1} shows the 3D phase portrait, highlighting the long-term transient chaotic dynamics of model \eqref{eq:03} in state space. This behavior is further validation for the presence of transient chaos displayed by the maximum  Lyapunov exponent in Figure \ref{fig:lyapanov1}.

\subsection{Results for model \ref{eq:insectide04}}
\begin{figure}[H]
\centering
\begin{subfigure}{.32\textwidth}
\centering
\includegraphics[width= 5.8cm, height=5cm]{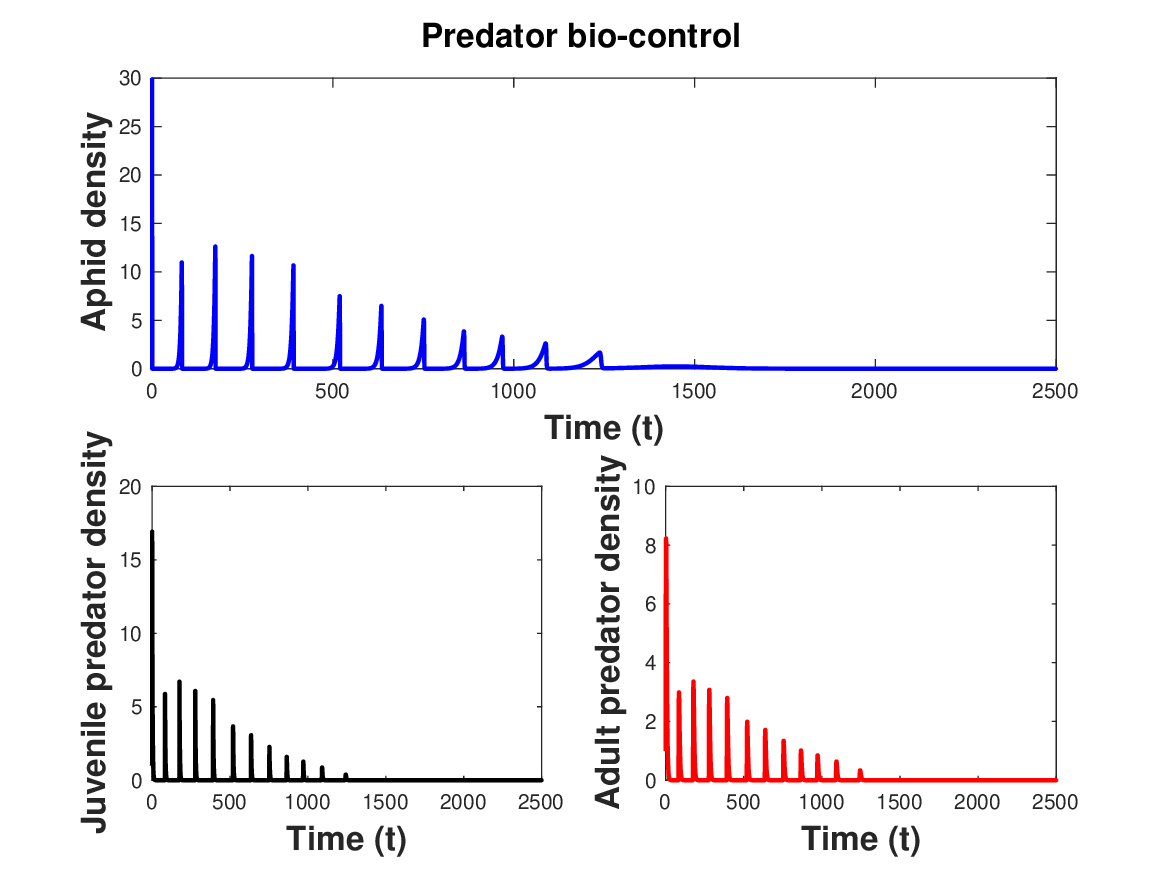}
\subcaption{No insecticide case}
\label{fig:pred9}
 \end{subfigure}
\begin{subfigure}{.32\textwidth}
\centering
\includegraphics[width=5.8cm, height=5cm]{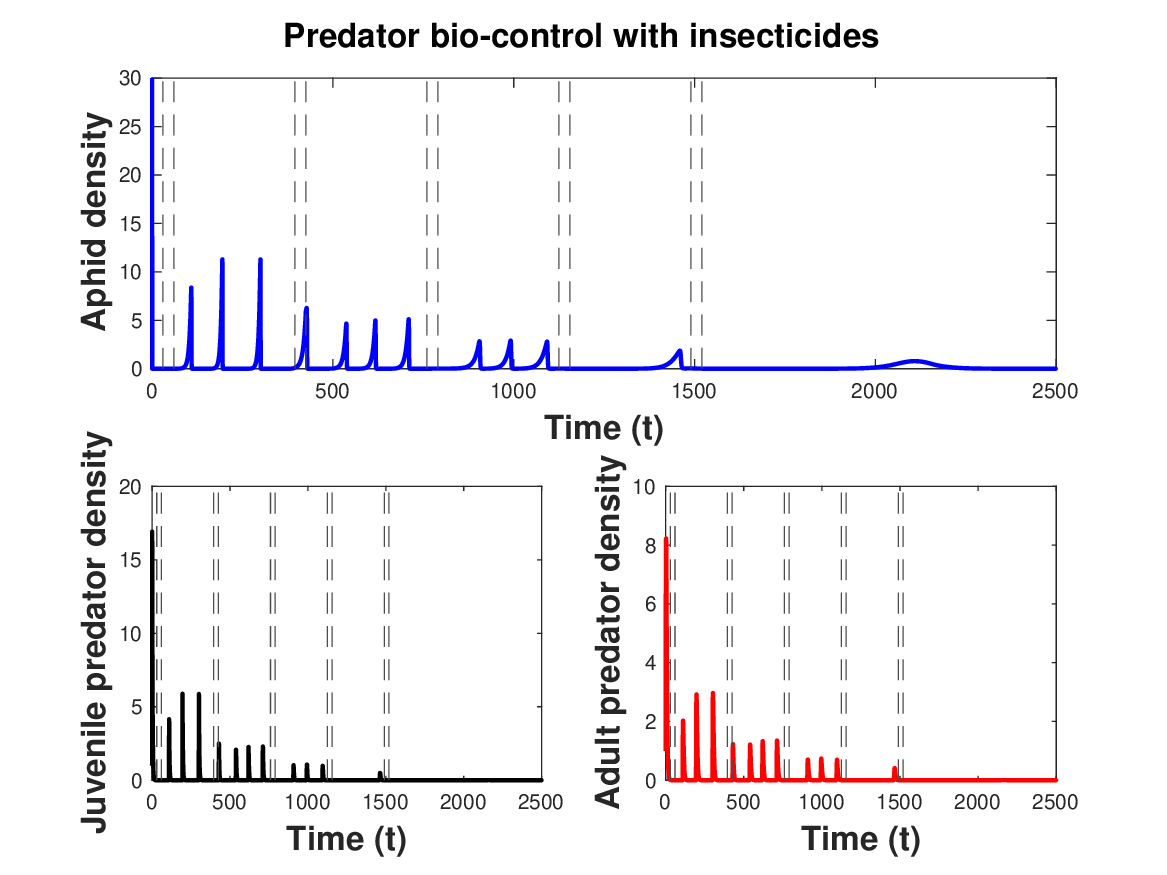}
\subcaption{Aphid \& predator dynamics}
\label{fig:aphid_inds}
  \end{subfigure}
    \hfill
  \begin{subfigure}{.32\textwidth}
  \centering
  \includegraphics[width= 5.8cm, height=5cm]{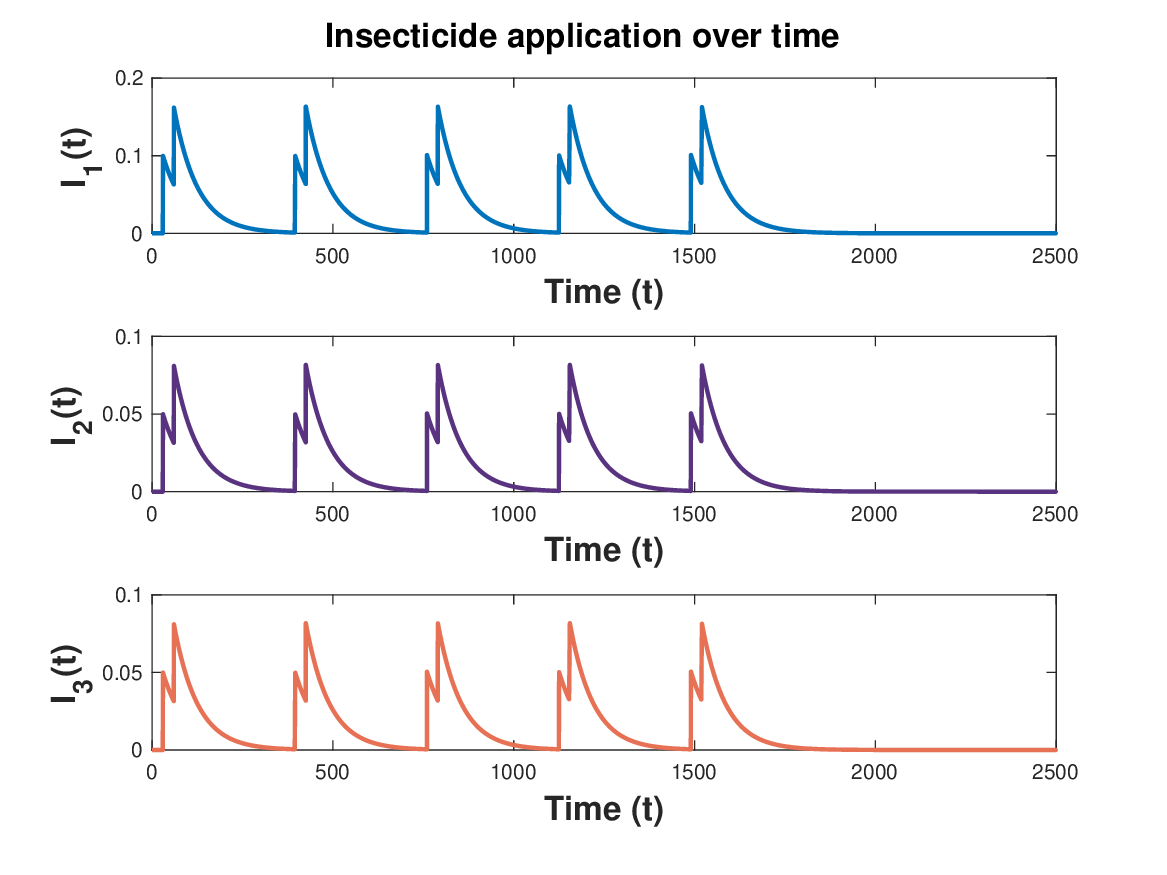}
  \subcaption{Re-application of insecticide}
\label{fig:ins_spray}
 \end{subfigure}
   \hfill
  
\caption{Time series plot for model \eqref{eq:insectide04}. In Figure \ref{fig:aphid_inds}, the top plot shows aphid population density over time, where the repeated application of insecticides reduces the number of peaks compared to no insecticide treatment. Without insecticide, the aphid population would die out early, but with insecticide, aphids experience smaller outbreaks over a longer period before eventually dying out. The insecticide was sprayed two times in the season, and this pattern continues for every season, giving the spray set = $[30, 60, 395, 425, 760, 790,1125,1155,1490,1520]$ reflected in \ref{fig:ins_spray}. 
The parameters used are $ a=0.0005, r=0.3, e_x=0.55, a_x=1.6281,t_x= 0.0260,\beta_j=0.4,\xi=0, \delta_a=0.3, c=\text{log}(2)/45, \delta_1=0.1,\delta_2=\delta_3=0.05$ with I.C.$=[0,30,1,1]$.}
\label{fig:pred_inset}
\end{figure}

\begin{figure}[H]
\centering
\begin{subfigure}{.32\textwidth}
\centering
\includegraphics[width= 5.8cm, height=5cm]{pred_insec_no_insec.eps}
\subcaption{No insecticide case}
\label{fig:pred_10}
 \end{subfigure}
\begin{subfigure}{.32\textwidth}
\centering
\includegraphics[width=5.8cm, height=5cm]{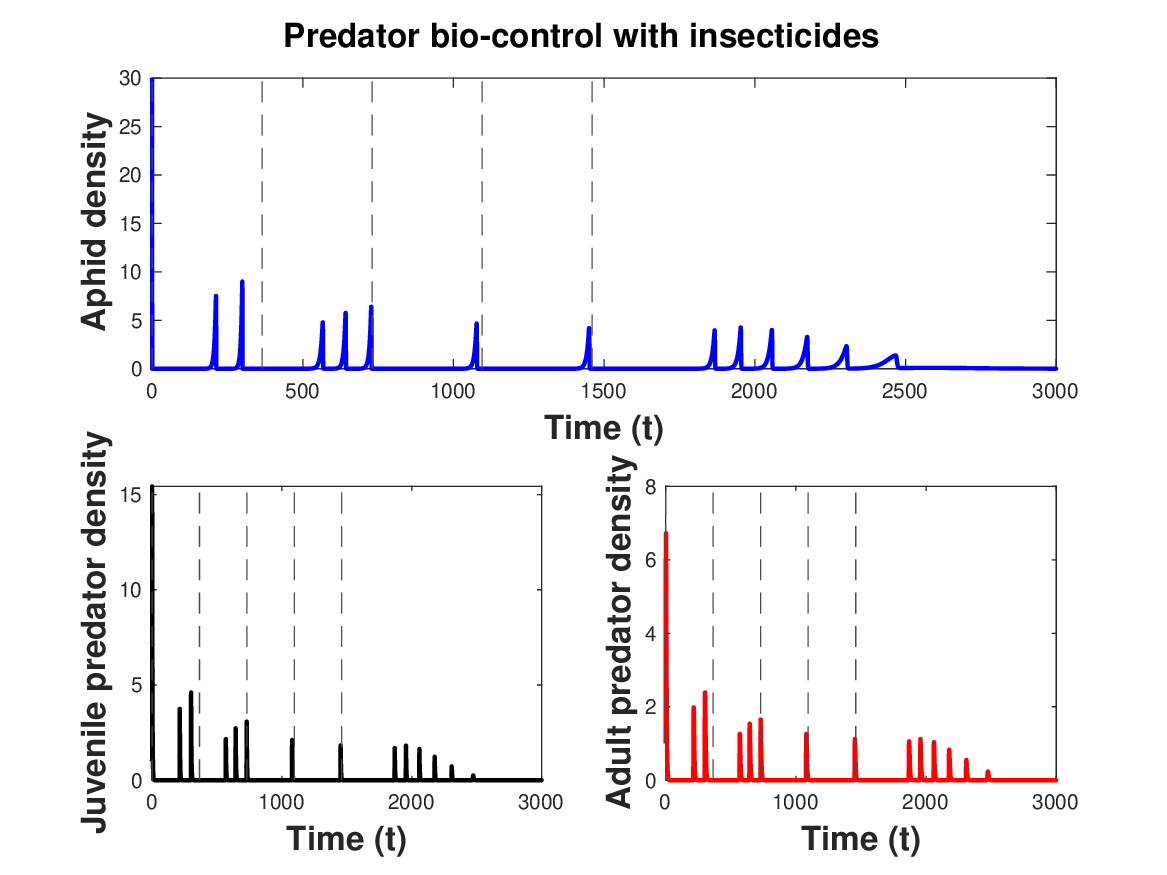}
\subcaption{Aphid \& predator dynamics}
\label{fig:annual_dynamics}
\end{subfigure}
\hfill
\begin{subfigure}{.32\textwidth}
\centering
\includegraphics[width= 5.8cm, height=5cm]{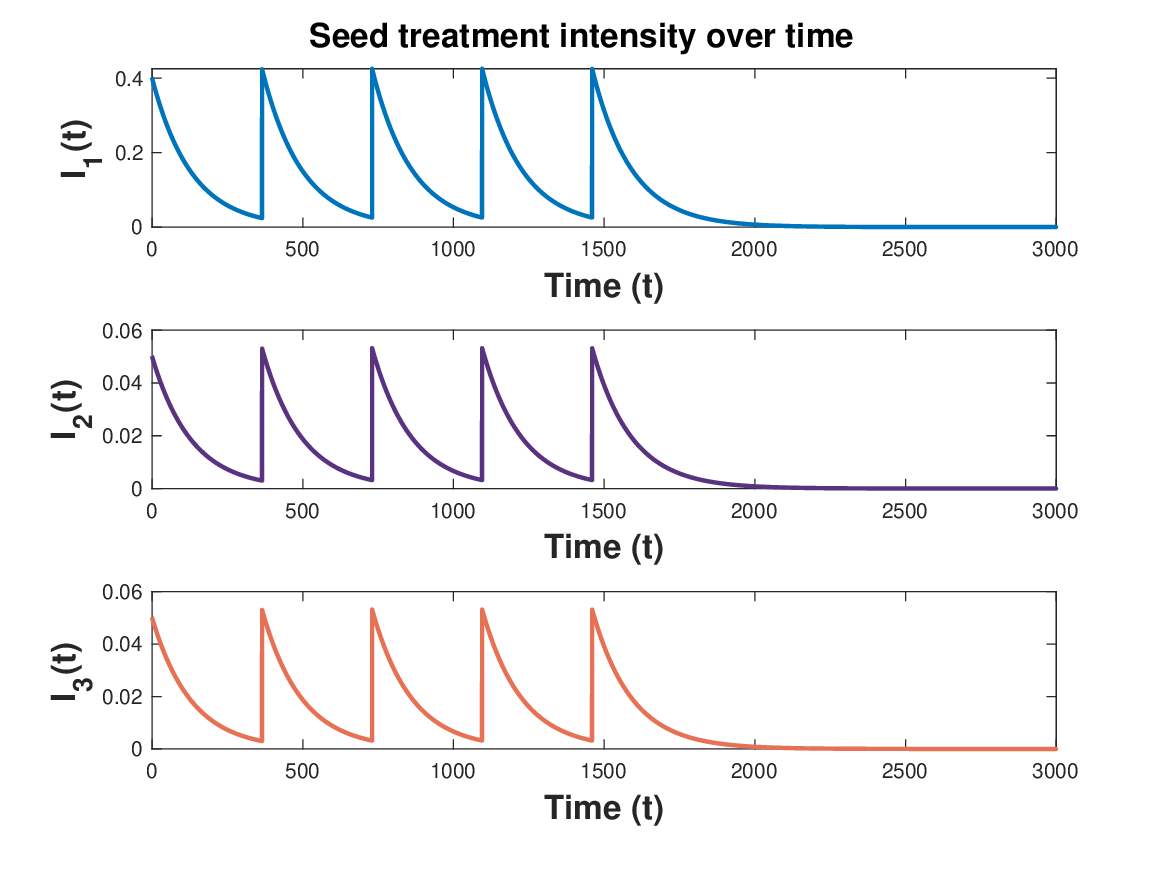}
\subcaption{Yearly insecticide application}
\label{fig:insec_one}
\end{subfigure}
\hfill    
\caption{Time series plot for model \eqref{eq:insectide04}. Figure \ref{fig:annual_dynamics} represents the population dynamics under seed treatment at the beginning of the season. The insecticide was applied once in the season, and this pattern continues for every season, giving the spray set = $[0,365,730,1095,1460,1825]$ reflected in \ref{fig:insec_one}. 
The parameters used are $ a=0.0005, r=0.3, e_x=0.55, a_x=1.6281,t_x= 0.0260,\beta_j=0.4,\xi=0, \delta_a=0.3, c=\text{log}(2)/90, \delta_1=0.4,\delta_2=\delta_3=0.05$ with I.C.$=[0,30,1,1]$.}
\label{fig:pred_annual_insect}
\end{figure}

We used simulations to check the impact of insecticides on the aphid population. Figure \ref{fig:pred_inset} explains the model \eqref{eq:insectide04} with a time series plot. In Figure \ref{fig:pred9} and Figure \ref{fig:pred_10}, no insecticide was applied, and we know from Figure \ref{fig:lyapanov1} that the system is chaotic for the chosen parameter set.  Figure \ref{fig:aphid_inds} represents when the insecticides are applied two times in a growing season at $t=30,60$  and in the same pattern for future years. It can be clearly seen that the use of insecticides has a negative effect on the aphid population. However, the use of insecticides appears to prolong the aphid's survival, delaying the time it takes for the population to go extinct.  Figure \ref{fig:ins_spray} represents the intensity of insecticide on aphids and predators over time. The timing of insecticide spraying is based on literature that shows spraying for soybean aphids after the R6 growth stage, will not improve yield (\cite{dean_hodgson_2024}). Seed treatments like neonicotinoids are another type of insecticide used to control  soybean aphids. Herein the seeds are chemically treated at the beginning of the season. Model \eqref{eq:insectide04} can represent this kind of treatment as well, by considering insecticide application once a year (see Figure \ref{fig:pred_annual_insect}).  The intensity/impact of the seed treatment on aphids is high in the initial phase and wears off with time (see Figure \ref{fig:insec_one}). The half-life of Neonicotinoids is taken to be 90 days, and it can be seen from the Figure  \ref{fig:annual_dynamics} that, because of high impact on aphids, the aphid population decreases, but not only the aphids survive longer, once the seed treatment is stopped, the outbreaks in aphid population occurs more frequently when compared to the case of foliar insecticide application (see figure \ref{fig:aphid_inds}). One possible reason could be the initial strong intensity in the case of seed treatments than multiple insecticidal sprays. 


\subsection{Results for model \ref{eq:05}}
\begin{figure}[H]
\centering
\includegraphics[width = 7cm, height=5cm]{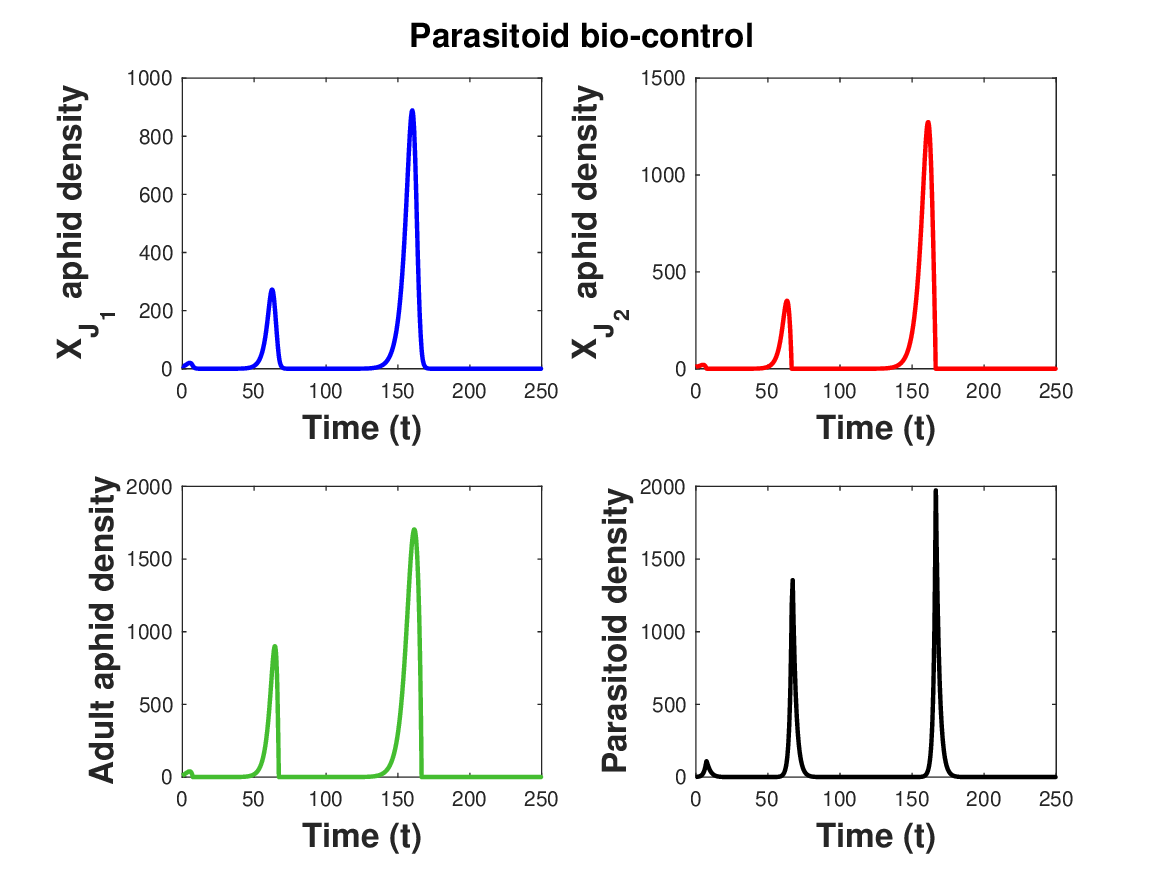 }
\caption{The time series plot shows the population dynamics of aphids under parasitoid attack for model \eqref{eq:05}. The four subplots represent the density changes over time for different aphid life stages (juvenile 1: top-left, juvenile 2: top-right, and adult aphids: bottom-left) and parasitoids: bottom-right. The parameters used are $ a=0.000005, r=0.65, \alpha=0.012,\beta_1=0.947,\beta_2=0.6124,\gamma_1=0.2767,\gamma_2=0.4671,\eta_1=0.5875,\eta_2=0.5667,\delta=0.45$ with I.C.$=[0,10,10,10,1]$.}
\label{fig:parasitoid1}
\end{figure}

The parameters are derived from data collected in the experimental setup described in section \eqref{Evaluating Parasitoid Preference for Different Aphid Instars}, and in particular from Table \ref{aphid_rates}. The attack rate and emergence rate used in simulations are based on the mean observed values, which are $\eta_1=0.5875, \eta_2=0.5667, \gamma_1=0.2767, \gamma_2=0.4671$. The maturation rates from one stage to another, i.e., $\beta_1$ and $\beta_2$, are calculated from Supplemental Table 
\ref{transfer rates}, 
which summarizes the number of aphids that entered and successfully exited a particular instar. Based on this table, $94.7 \%$ of the aphids that were tracked successfully moved from the $1^{\text{st}}$ / $2^{\text{nd}}$ instar age group into the $3^{\text{rd}}$ / $4^{\text{th}}$ instar age group, giving us $\beta_1$. And $88.9\%$ of the aphids successfully moved from the $3^{\text{rd}}$ / $4^{\text{th}}$ instar age group into the adult life stage, which gives us modified $\beta_2$ after including the attack rate from parasitoid on the $j_2$ stage. Figure \ref{fig:parasitoid1} represents the time series plot for model \eqref{eq:05}, showing that multiple peaks can occur across all aphid stages, followed by a delayed response and peaks in the parasitoid population. 

\subsection{Results for model \ref{eq:09}}

\begin{figure}[H]
\begin{subfigure}{.48\textwidth}
\centering
\includegraphics[width=7cm, height=5cm]{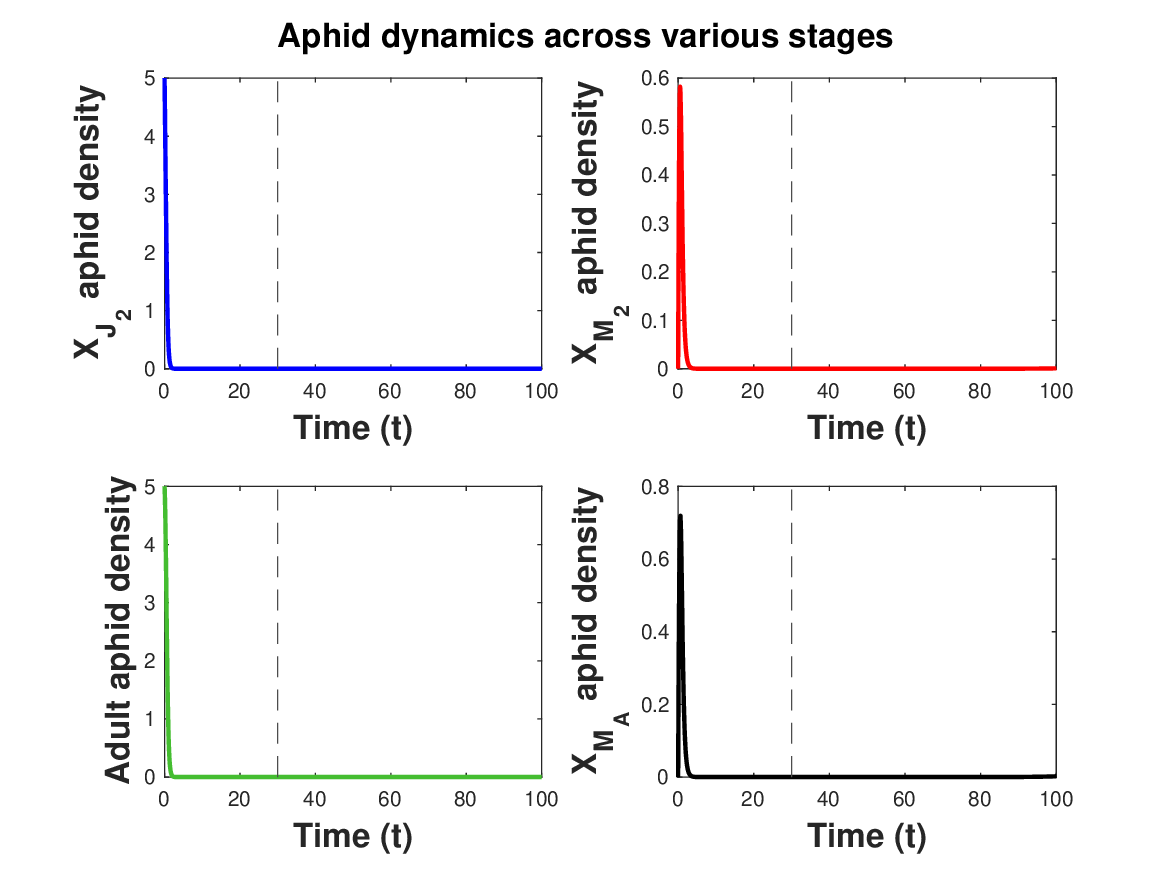}
\subcaption{Aphid population dynamics}
\label{fig:aphid_comb}
  \end{subfigure}
  \begin{subfigure}{.48\textwidth}
  \centering
  \includegraphics[width= 7cm, height=5cm]{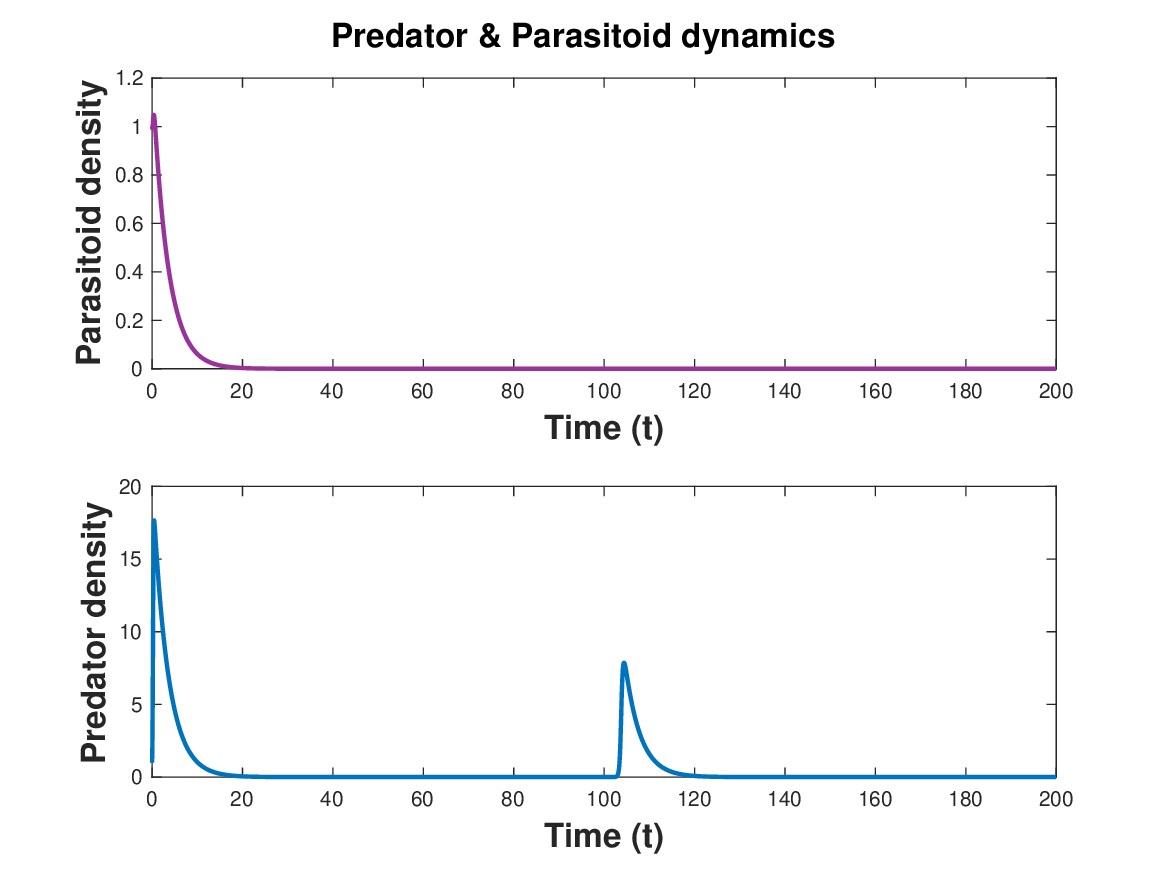}
  \subcaption{Natural enemy dynamics}
\label{fig:pred_para_comb}
 \end{subfigure}
\caption{The time series plot shows the population dynamics of aphids, parasitoids, and predators for model \eqref{eq:09}. In Figure  \ref{fig:aphid_comb}, the four subplots represent the density changes over time for different aphid life stages (juvenile: top-left, mummified juvenile: top-right, adult aphids: bottom-left, and mummified adult: bottom-right). Figure \ref{fig:pred_para_comb} represents the parasitoid (top subplot) and predator (bottom subplot) population dynamics over time.  From the simulation, it can be seen that peaks later in the season can occur in aphids, followed by a delayed response and peak in the predator population, while the parasitoid population crashes early in the season, showing the impact of intraguild predation. The parameters used are $ a=0.000005, r=0.3, \alpha=0.012,\beta_1=0.6,\beta_2=\beta_3=0.45,\gamma_1=0.2,\gamma_2=0.38,\eta_1=0.5875,\eta_2=0.5667,\epsilon_1=\epsilon_2=0.55,q_1=q_2=0.7,a_1=a_2=1.6281,t_1=t_2=0.0260,\delta_p=\delta_j=0.3,\delta_1=\delta_2=\delta_3=0,c=0$ with I.C.$=[0,15, 0, 15, 0, 1, 1]$.}

\label{fig:parasitoid_predator_comb}
\end{figure}

 For the model \eqref{eq:09}, the parameters were chosen to highlight the influence of intraguild predation. The parameter values associated with the predator chosen from section \eqref{pred_section_1}, and biologically reasonable values were chosen for parasitoid-related parameters. The probabilities $q_1,q_2$ were set high,  as studies suggest that predators prefer non-parasitized aphids over the parasitized ones (\cite{xue2012intraguild,fu2017intraguild}). Figure \ref{fig:parasitoid_predator_comb} represents the population dynamics of aphids with predators and parasitoids when no insecticide is used on any population group. This figure shows that a strategy of using parasitoids, where additional stage structure is assumed (i.e., mummified aphids) in conjunction with predators and insecticides, stabilizes the system \eqref{eq:03} when compared to Figure \ref{fig:pred7}, where only predators are used. The parameters related to predator dynamics are kept the same as in Figure \ref{fig:pred7}, where transient chaotic dynamics were present. 

\subsection{Long term dynamics}

\begin{figure}[H]
\centering
\begin{subfigure}{.32\textwidth}
\centering
\includegraphics[width= 5.8cm, height=5cm]{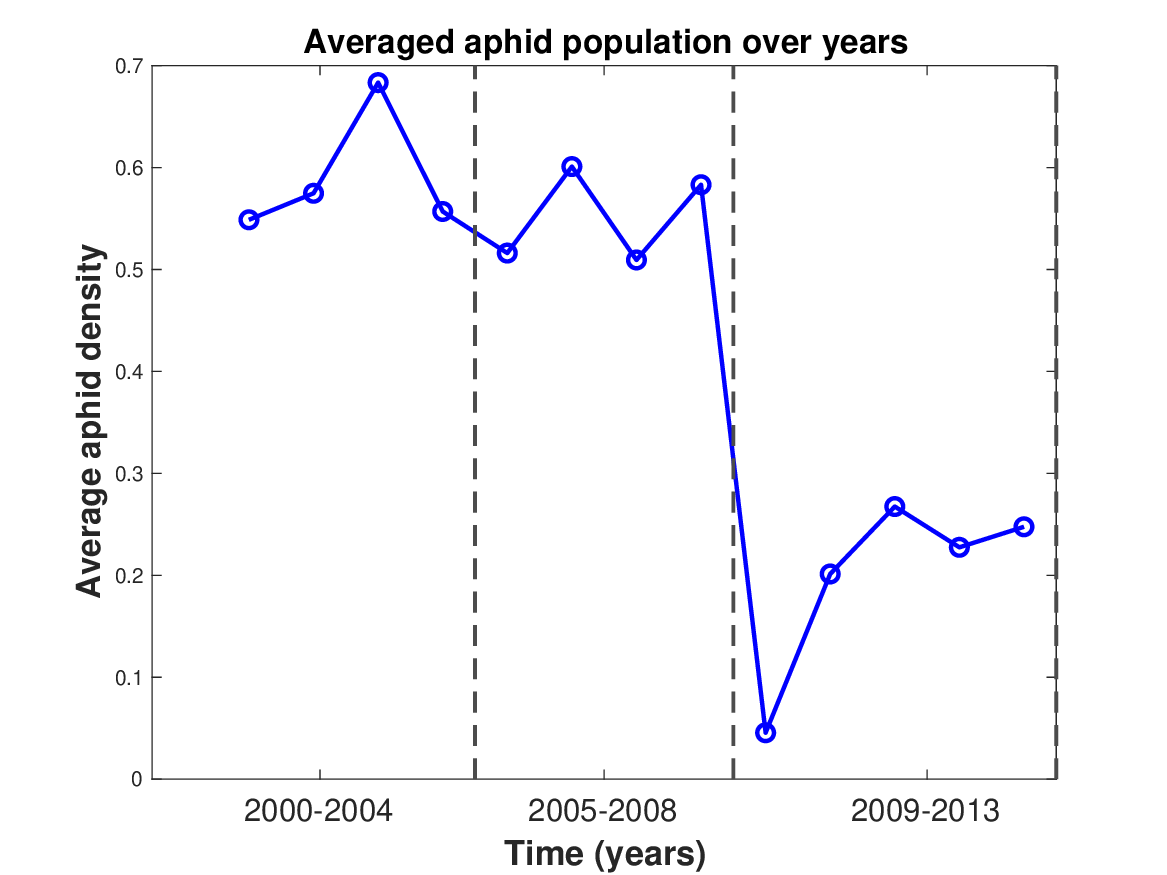}
\subcaption{}
\label{combined aphid}
\end{subfigure}
\begin{subfigure}{.32\textwidth}
\centering
    \includegraphics[width= 5.8cm, height=5cm]{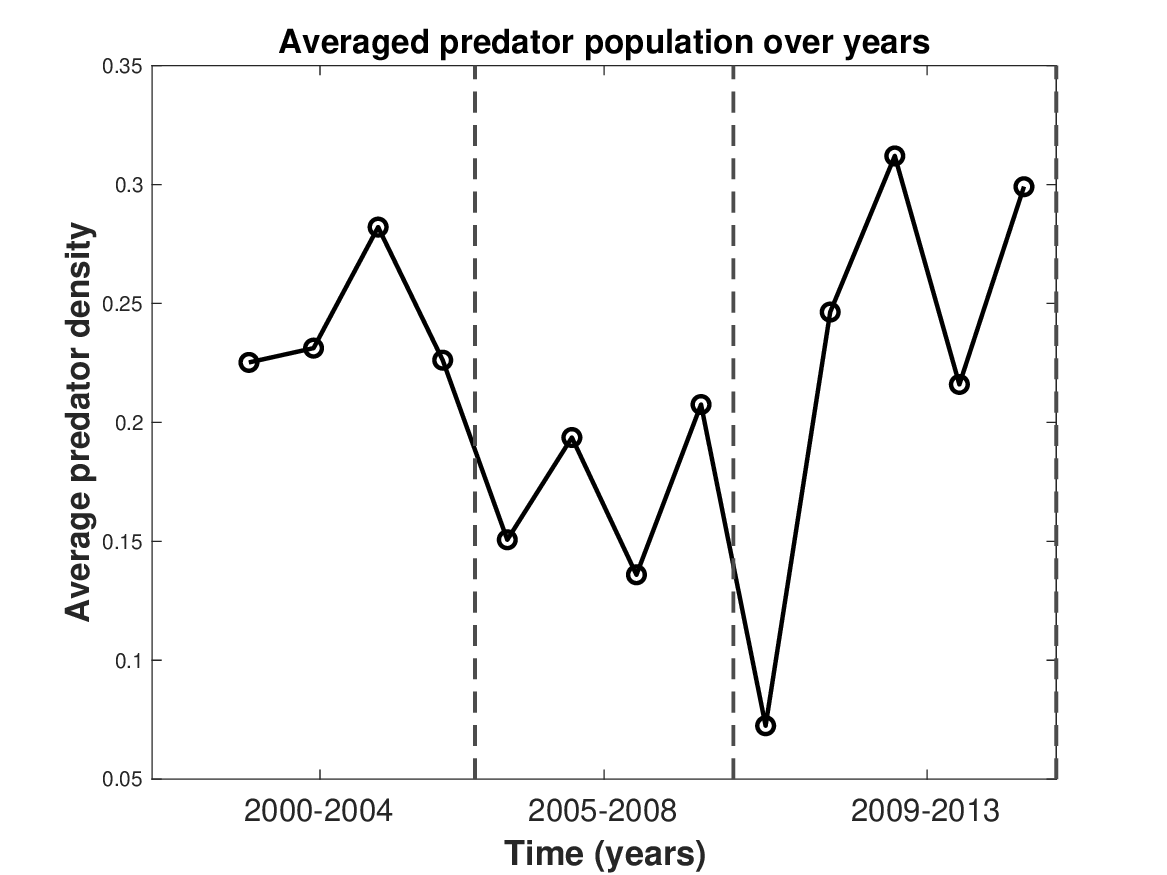}
    \subcaption{}
    \label{combined pred}
\end{subfigure}
\begin{subfigure}{.32\textwidth}
\centering
\includegraphics[width= 5.8cm, height=5cm]{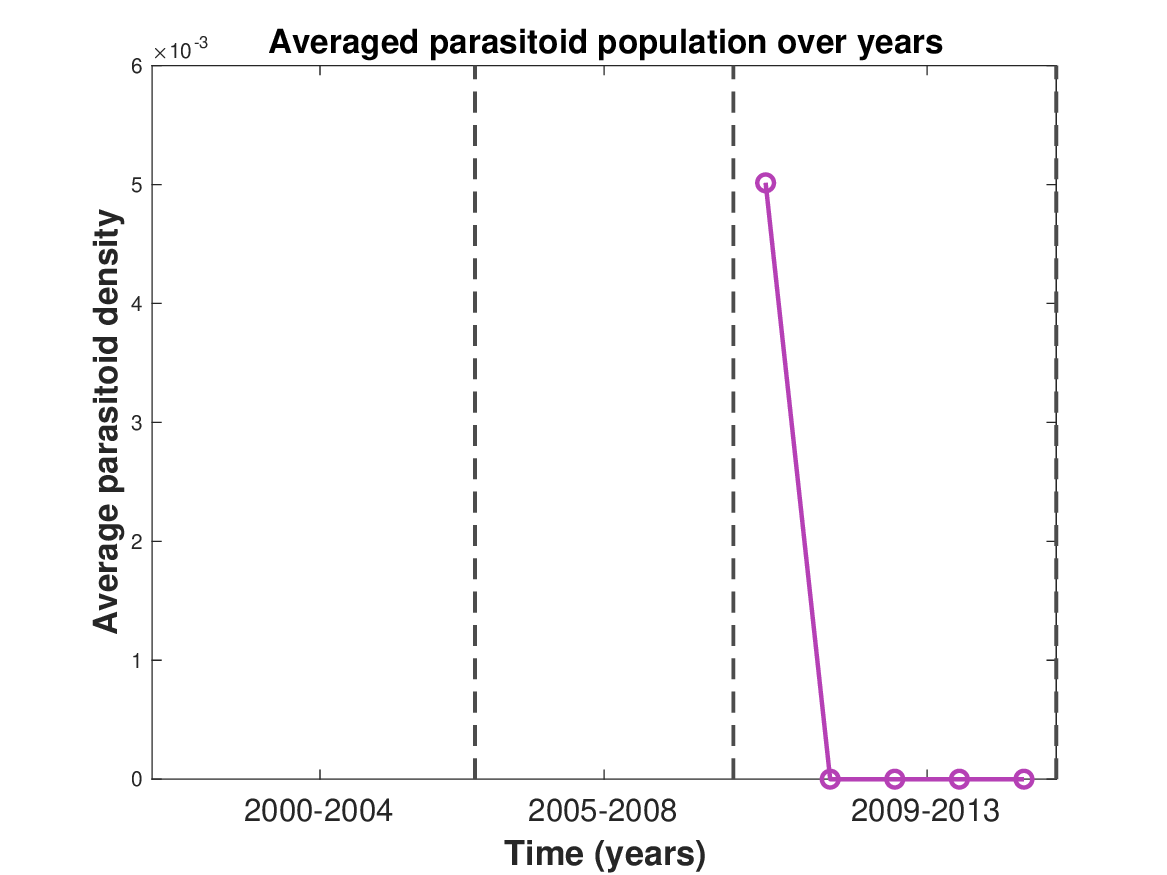}
\subcaption{}
\label{combined parasitoid}
\end{subfigure}
\caption{Population densities for aphids, predators, and parasitoids over 13 years $(2000-2013)$. The time series plot is divided into three different phases corresponding to three different models representing the population dynamics in the given period. The initial condition in the first phase was $[0,30,1,1]$, and the initial condition for the model \eqref{eq:insectide04} in phase 2 is taken from the end values of phase 1 variables with some survival probability, to begin the next phase with a reasonable population. The model in phase 3 is different from the previous two models, as there is a stage structure in aphids because of the parasitoid introduction. Hence, the initial condition is chosen carefully with survival probability from the end values of the phase 2 variables. The initial predator population in phase 3 is determined by adding the end values of adult and juvenile predators from phase 2. The initial parasitoid density is chosen to maintain a ratio of approximately one parasitoid for every $15$ aphids. The parameters used are $ a=0.000005, r=0.3, \alpha=0.012,\beta_1=0.6,\beta_2=\beta_3=0.45,\gamma_1=0.2,\gamma_2=0.38,\eta_1=0.5875,\eta_2=0.5667,\epsilon_1=\epsilon_2=\epsilon_x=0.55,q_1=q_2=0.6,a_x=a_1=a_2=1.6281,t_x=t_1=t_2=0.0260,\beta_j=0.4,\delta_a=\delta_p=\delta_j=0.3,c=\text{log}(2)/45,\delta_1=0.15,\delta_2=\delta_3=0.05$.}

\label{combined_populations}
\end{figure}

We also simulated long-term aphid dynamics across three different phases. Phase 1 $(2000-2004)$: the period when only predators were originally present and this situation can be modeled by \eqref{eq:03}, Phase 2 $(2005-2008)$: the phase when the use of insecticides increased rapidly and this is described by model \eqref{eq:insectide04}. Phase 3 $(2009-2013)$: the period during which the parasitoids, as another bio-control agent, were introduced along with the predator and insecticide treatment, and this is modeled by \eqref{eq:09}. Figure \ref{combined_populations} represents the average aphid population along with average predator and parasitoid populations for a period of $13 \  \text{years} \ (2000-2013)$. The corresponding time series plot for figure \ref{combined_populations} is supplemental figure 
\ref{fig:combined_ts}. 
The insecticide treatment started in the year $ 2005$ and was done twice in a year, and then followed the same pattern till $2013$, refer to supplemental figure 
\ref{fig:combined_insec}. 
The averaged predator plot qualitatively captures the population density trend of the Asian lady beetle as shown in Figure 1 of (\cite{bahlai2015shifts}).

\subsection{Simulations for Economic Injury Level (EIL) and Economic Threshold (ET)}

In this section, we present various simulations for models \eqref{eq:0}, \eqref{eq:insectide04}
and \eqref{eq:09}. The main idea is to understand the effect of natural enemy control and the use of insecticides in suppressing soybean aphids below economically important levels. Particularly to understand the impact of predator and parasitoid control on the Economic Injury Level (EIL) and Economic threshold (ET).  

\vspace{0.2cm}
The common parameters were fixed for all three different models, where there is no control  \eqref{eq:0} corresponds to the single species aphid model, when there is predator control with insecticide treatment \eqref{eq:insectide04}, and when parasitoid is introduced with existing control strategies \eqref{eq:09}. 
First, we ran the time series for all these models for a single season $t=100$ days, to note the peak aphid population, the time when aphids surpass the economic threshold. We also looked at the average aphid population as well as the populations of the predator and parasitoid over time in a single growing season.

\begin{figure}[H]
 \begin{subfigure}{.48\textwidth}
\centering
 \includegraphics[width=7cm, height=5cm]{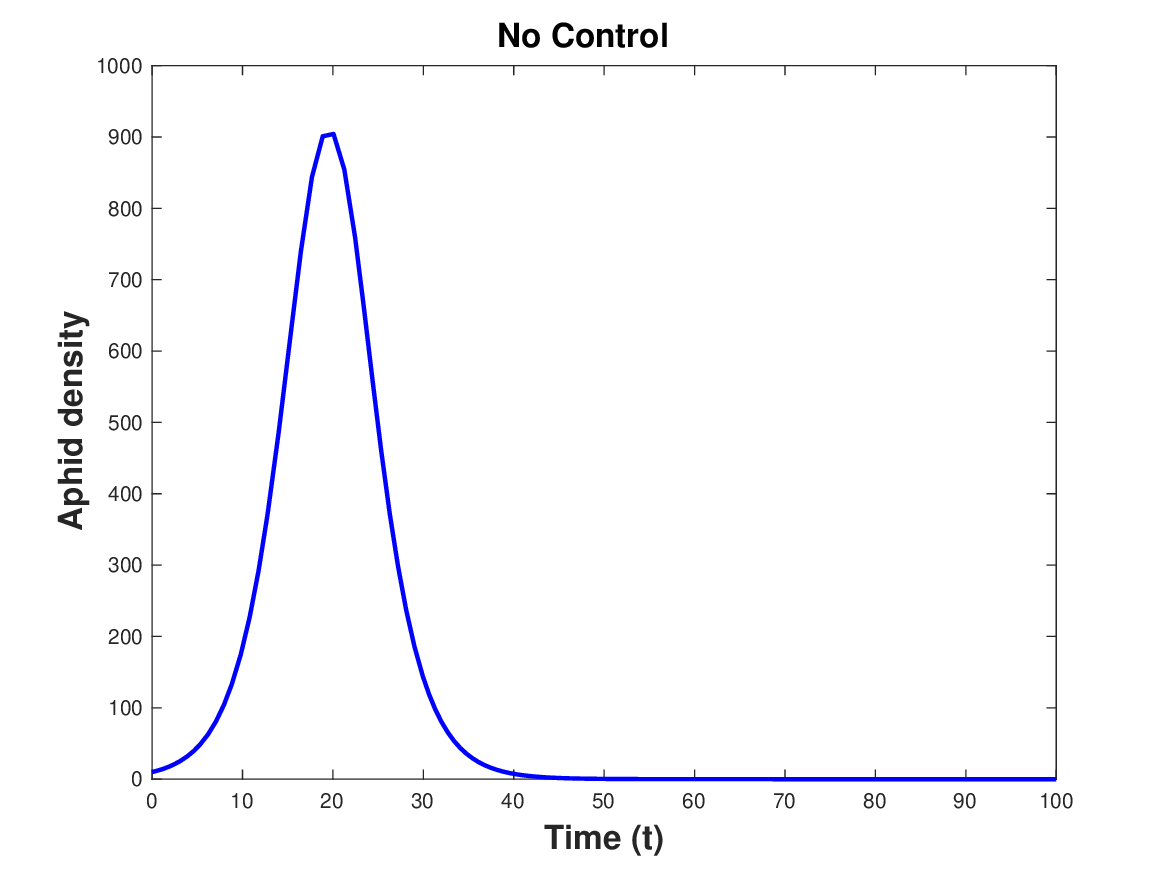}
 \subcaption{ }\label{fig:first_no_average_para}
  \end{subfigure}
  \begin{subfigure}{.48\textwidth}
  \centering
  \includegraphics[width= 7cm, height=5cm]{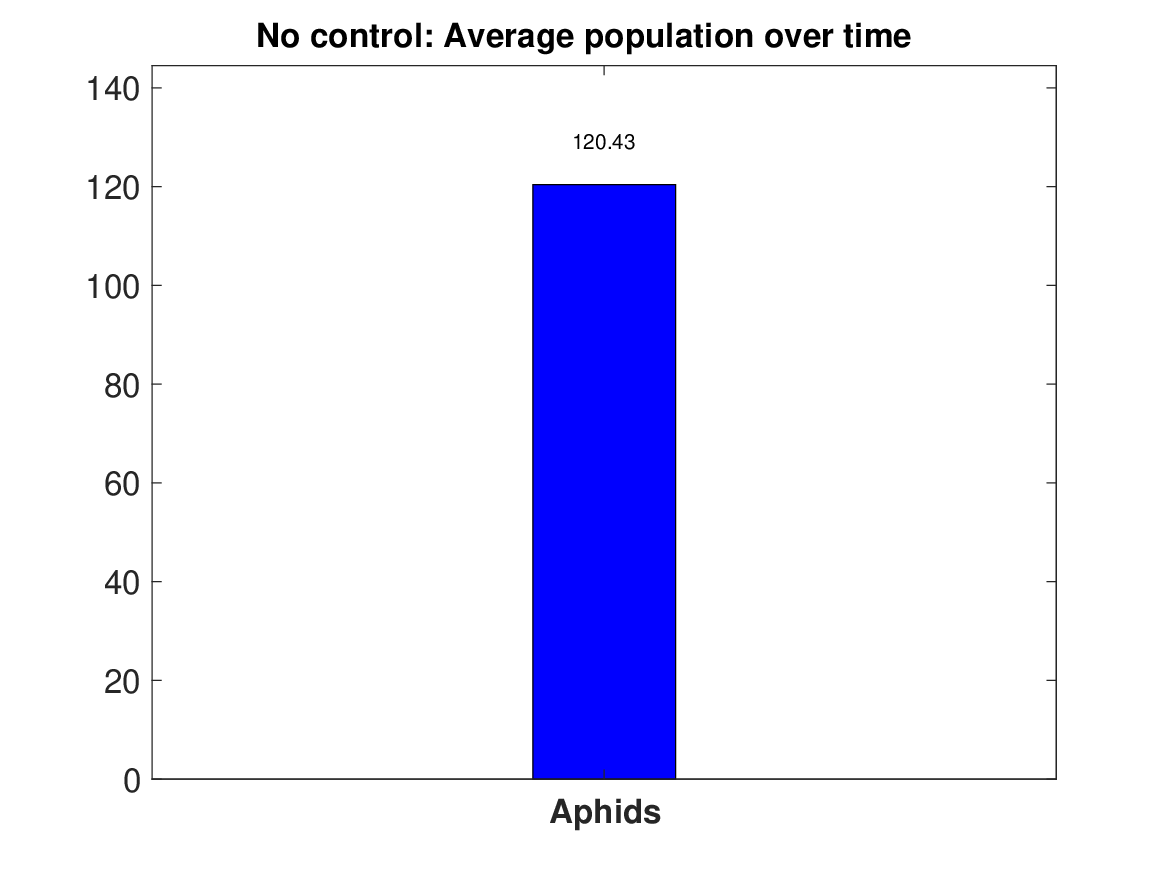}
  \subcaption{}
\label{fig:second_no_average_para}
 \end{subfigure}
 
 \caption{These subplots show \ref{fig:first_no_average_para} changes in aphid population density over time and \ref{fig:second_no_average_para} the average aphid population over a single field season for model \eqref{eq:0}, which corresponds to a single-species aphid model with no control. The scaling parameter and aphid growth rate are fixed at $a= 5 \cross 10^{-5}, r=0.3$ with initial density as $[0,10]$. It can be seen with no control, aphid grows rapidly and crosses ET on day $11$ and continues to grow and reach the injury level earlier in the season.}
 
\label{plot_average_no_control}
\end{figure}

\begin{figure}[H]
  \begin{subfigure}{.48\textwidth}
\centering
  \includegraphics[width=7cm, height=5cm]{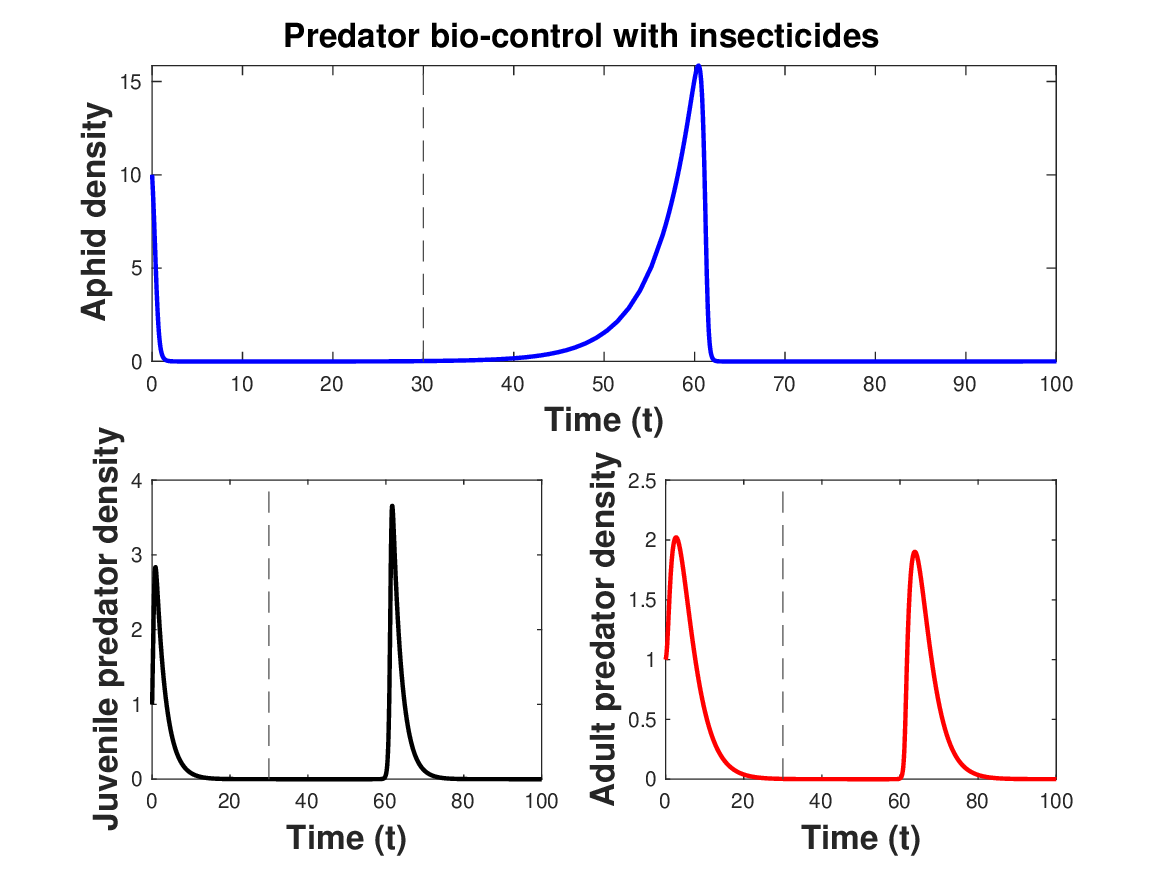}
 \subcaption{ }
 \label{pred_ts_1}
  \end{subfigure}
  \begin{subfigure}{.48\textwidth}
  \centering
  \includegraphics[width= 7cm, height=5cm]{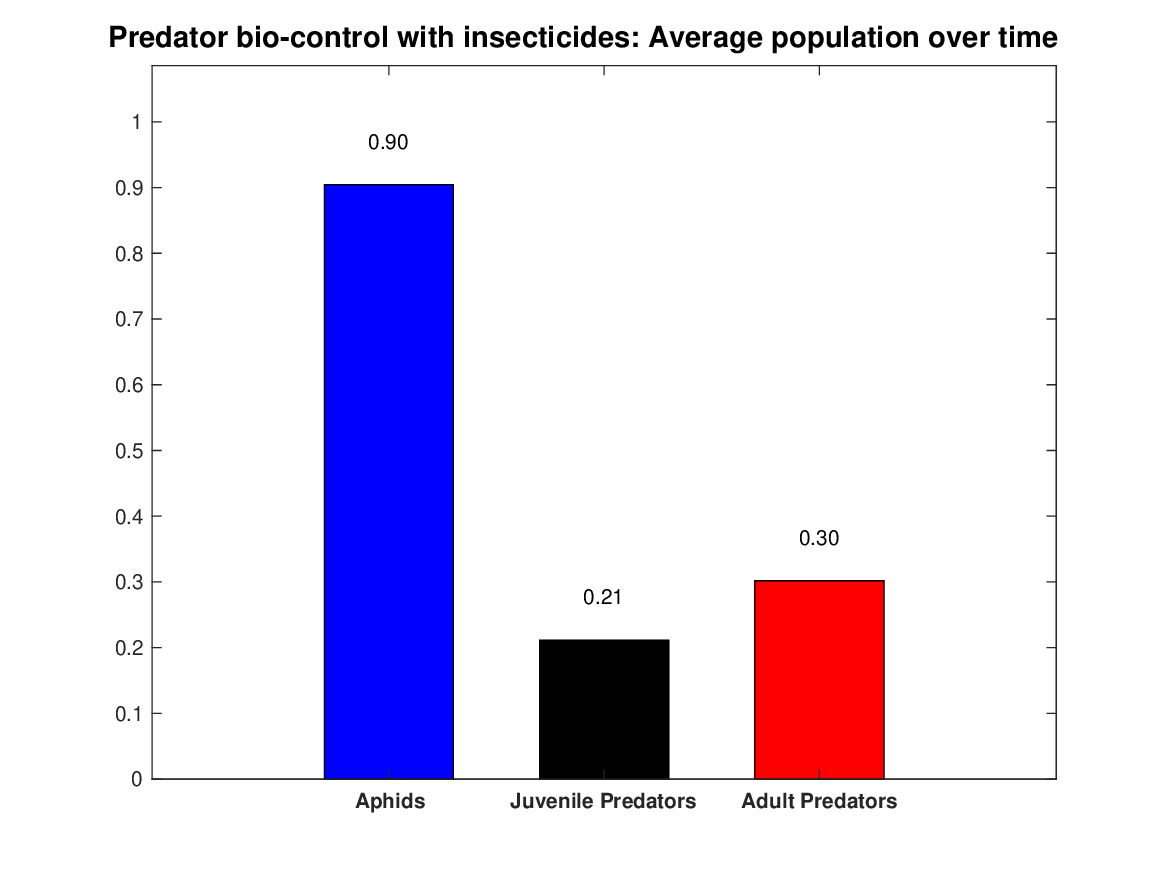}
  \subcaption{}
\label{pred_ts_2}
 \end{subfigure}

 \caption{ These subplots show \ref{pred_ts_1} changes in aphid and predator population density over time and \ref{pred_ts_2} the average aphid and predator population over a single field season for model \eqref{eq:insectide04}, which represents the predator biocontrol with insecticide application. The blue bar chart represents the average aphid population, while the black and red bar chart shows the average populations of juvenile and adult predators, respectively. The insecticide was sprayed once at $t=30$, which delayed the peak in aphid population, and aphid count never crossed the economic threshold in the season. The parameters used are $ a= 5 \cross 10^{-5}, r=0.3, e_x=0.25, a_x=1.6281,t_x= 0.0260,\beta_j=0.4,\xi=0, \delta_a=0.3, c=\text{log}(2)/45, \delta_1=0.1,\delta_2=\delta_3=0.05$ with I.C.$=[0,10,1,1]$.
 }
\label{plot_average_pred_control}
 \end{figure}

\begin{figure}[H]
  \begin{subfigure}{.32\textwidth}
\centering
  \includegraphics[width=5.8cm, height=5cm]{el_aphid_3.eps}
 \subcaption{ }
 \label{fig:first_entry_average_para}
  \end{subfigure}
  \begin{subfigure}{.32\textwidth}
  \centering
  \includegraphics[width= 5.8cm, height=5cm]{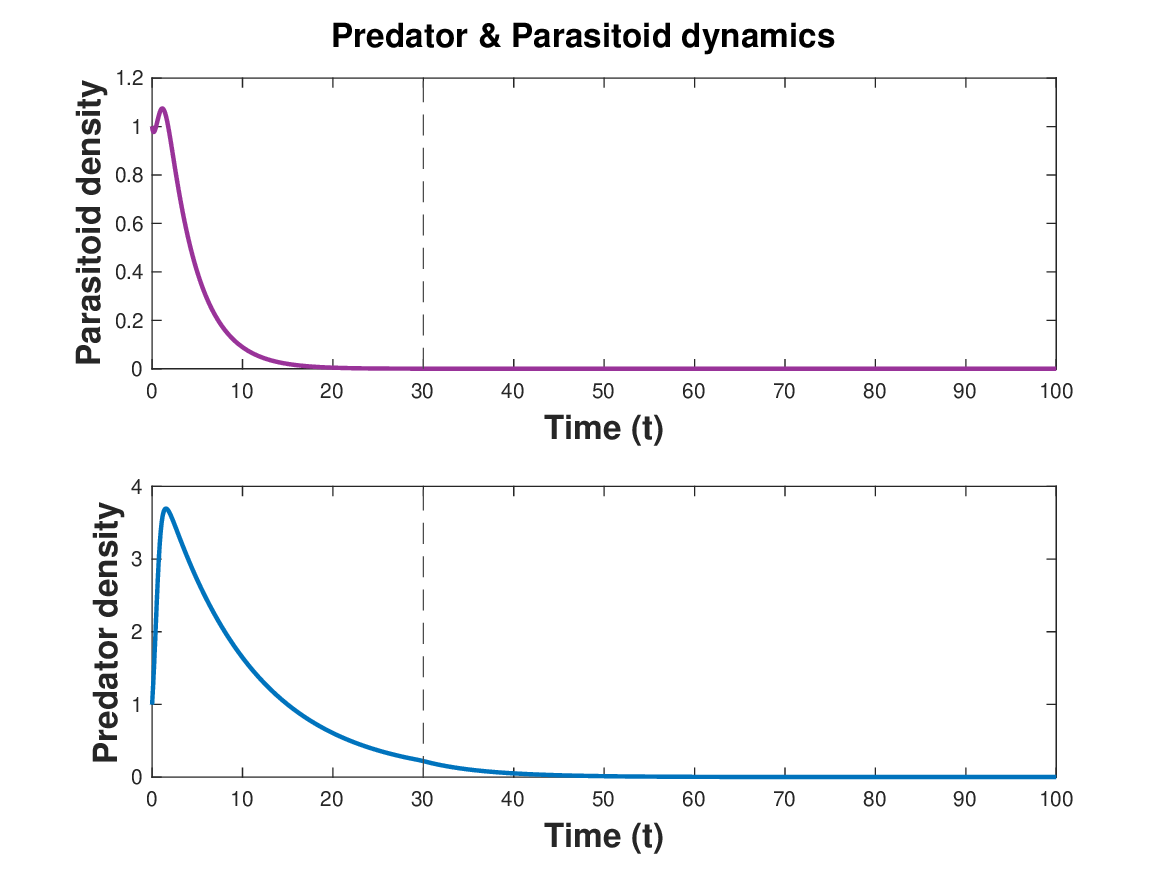}
  \subcaption{}
\label{fig:second_entry_average_para}
 \end{subfigure}
  \begin{subfigure}{.32\textwidth}
  \centering
  \includegraphics[width= 5.8cm, height=5cm]{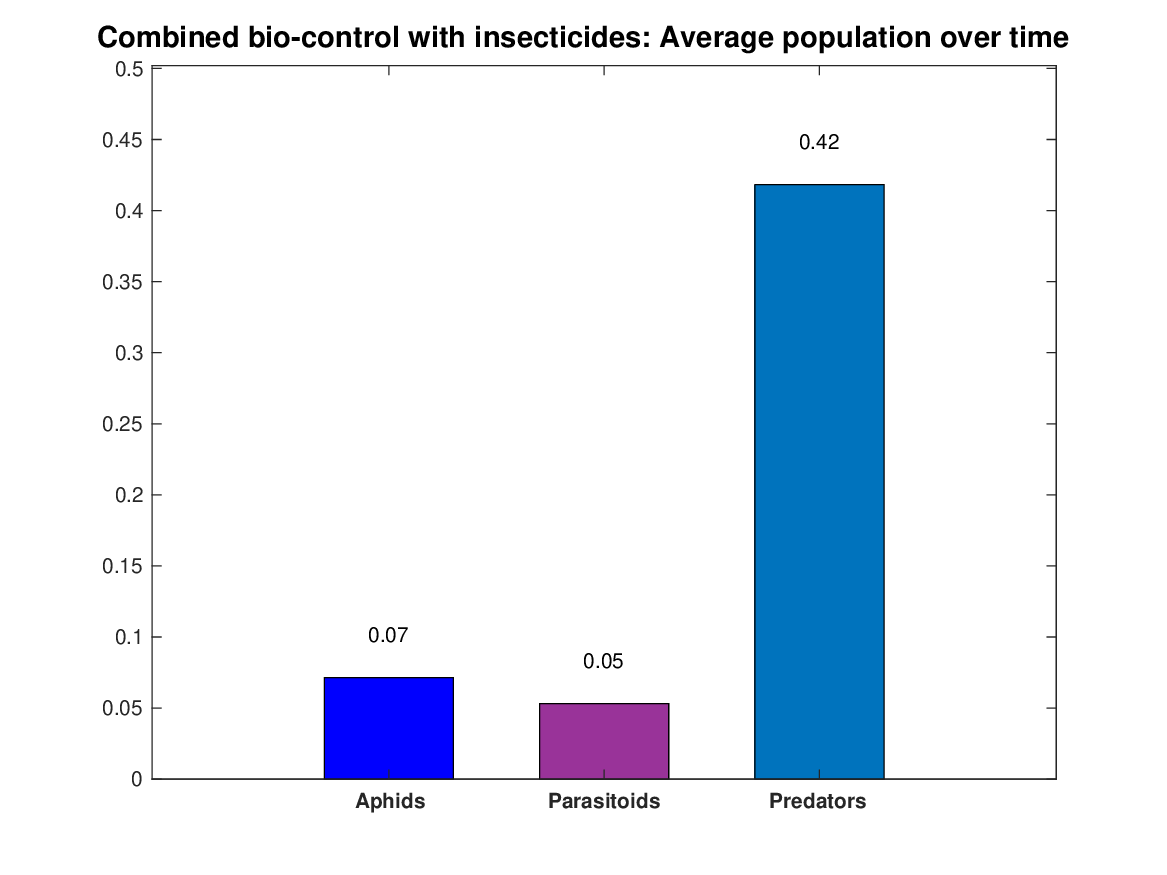}
  \subcaption{}
\label{fig:third_entry_average_para}
 \end{subfigure}
 \caption{These subplots shows \ref{fig:first_entry_average_para} changes in aphid population density across various stages over time (juvenile: top-left, mummified juvenile: top-right, adult aphids: bottom-left, and mummified adult: bottom-right), \ref{fig:second_entry_average_para} changes in predator and parasitoid population density over time and, \ref{fig:third_entry_average_para} shows the average aphid, parasitoid and predator population over a single field season \eqref{eq:09}, which accounts for the predator-parasitoid coexistence with insecticide treatment control strategy. The bar chart represents the average population over time as follows: blue for the aphids across all the juvenile and mummified stages, violet for the parasitoid population, and light blue for the predator population. The insecticide was sprayed once at $t=30$. It can be seen that the combined strategy suppressed the aphid population across all stages very early in the season, and the population never reached the economic threshold.  The parameters used are $ a= 5 \cross 10^{-5}, r=0.3, \alpha=0.012,\beta_1=0.6,\beta_2=\beta_3=0.45,\gamma_1=0.2,\gamma_2=0.38,\eta_1=0.5875,\eta_2=0.5667,\epsilon_1=\epsilon_2=0.25,q_1=q_2=0.6,a_1=a_2=1.6281,t_1=t_2=0.0260,\delta_p=0.1,\delta_j=0.3,\delta_1=0.1,\delta_2=\delta_3=0.05,c=\text{log}(2)/45$ with I.C.$=[0,5, 0, 5, 0, 1, 1]$.} 
\label{plot_average_para_control}
 \end{figure}

\vspace{0.2cm}
Initially, the parameter set is chosen for model \eqref{eq:0} for which the peak population of soybean aphids surpasses the ET level ($250$ aphids per plant) (\cite{ragsdale2011ecology}); see Figures \ref{fig:first_no_average_para} and  \ref{fig:second_no_average_para} shows the average aphid population over the season.   Then, keeping the same common parameter $a,r$, we run simulations and note the above details for model \eqref{eq:insectide04} and  \eqref{eq:09} as well (see figure \ref{plot_average_pred_control} and \ref{plot_average_para_control}). All other parameters come from the experimental setup or are biologically relevant values, as mentioned earlier. This approach helps us to evaluate whether a predator or parasitoid, along with chemical treatments, can (i) reduce the average population, and (ii) to what extent is this reduction when compared to the scenario without control (see table \ref{table:et_eil_levels}). However, it is important to note here that it requires a very long time (often multiple seasons) for bio-control agents to eradicate the aphids. The insecticide was sprayed for two years, and then aphids were attacked by natural enemies, as we had already noticed that the spraying of insecticide prolongs aphid survival (see figure \ref{fig:pred_inset}, \ref{fig:pred_annual_insect}). The criteria to stop the time series simulation are based on when the aphid population starts stabilizing and drops below a threshold value of $10^{-6}$ (see supplemental figures 
\ref{supp_plot_average_no_control}, \ref{supp_plot_average_pred_control}, and \ref{supp_plot_average_para_control}) 
for the respective time series and averaged population plot.  We consider ET or EIL based on a single time point as $t=100$ from an Integrated Pest Management (IPM) perspective, reflecting practical pest management strategies rather than relying on long-term eradication dynamics. 

\begin{singlespace}
\begin{table}[H]
 \centering
\begin{tabular}{|l|c|c|c|}
\hline
\textbf{Model} & \textbf{Peak Aphid Population } & \textbf{ET exceeded on} & \textbf{Average Aphid Population} \\
\hline
\eqref{eq:0}  & 904.1999 (on Day 20)  & Day 11 &  120.4250\\
 \hline
\eqref{eq:insectide04}& 15.8507  (on Day 60)  & No & 0.904532 \\
 \hline
\eqref{eq:09} & 10 (on Day 0)  & No & 0.071389 \\
\hline
\end{tabular}
\caption{Comparison of aphid population dynamics with respect to different bio-control strategies, showing peak population, day ET was crossed, average aphid population over a single season. The peak aphid population for model \eqref{eq:09} is calculated by combining the aphid population across stages and then finding the maximum, which never exceeds the combined initial population density.}
\label{table:et_eil_levels}
\end{table}
\end{singlespace}

\section{Discussion \& Conclusion}
The models presented in the manuscript simulate aphid dynamics with predators and parasitoids in multiple scenarios, and there are several important implications of our results. These models advance our capacity to simulate aphid natural enemy interactions, allowing us to explore biocontrol efficacy under various complex scenarios. Overall, the results from our modeling simulations suggest that natural enemies agents do have the capacity to suppress aphid populations below economically damaging levels under certain conditions. In particular, the most effective aphid suppression seems to occur in combined, top-down biocontrol models where both predators and parasitoids are present. The control is further enhanced if insecticides are used in conjunction with bio-control.

We introduce a top-down biocontrol model, where the aphid dynamics are modeled via (\cite{kindlmann2010modelling}). However, our model includes a numerical response for the predator. This is motivated empirically as the predator used for our experiments, the lacewing, has a life cycle of $\sim 27$ days, which yields a minimum of $3-4$ generations over the summer season if generations are non-overlapping; with overlap, the total number of seasonal generations is likely higher. This model incorporates age structure in the predator since only the larva stage feeds on the aphid.  
The predator-only bio-control model predicts transient chaotic dynamics. This is possibly the first evidence of such dynamics in bio-control models, where the dynamics of the pest is of boom-bust type. The chaos is quantified via a computation of the Lyapunov exponent with respect to the scaling parameter $a$. 

Further, we model the use of insecticides along with predators in two approaches: first, using foliar application twice in the growing season, and secondly, the insecticide is applied once only at the beginning of the season with strong intensity and wears off over time to mimic the seed treatment(such as use of neonicotinoid).  
The insecticide spraying does not stabilize the system unless applied in large quantities, and it prolongs the survivability of aphids, with later peaks appearing upon insecticide use - this could be interpreted as insect resistance developing. We also model neonicotinoid use with its half-life parametrized from the literature, which models a decreasing effect of the insecticide over time, but does not stabilize the system. This is similar to the existing pest management adopted in $2005-2008$, when insecticide use was increased. We also introduce a parasitoidal age-structured top-down biocontrol model, which also has structure in the pest (aphid) because of the parasitoids' preference for different life stages. This model does not exhibit chaotic dynamics, but exhibits multiple population peaks, which can vary in amplitude.   

Next, we also introduce a combined top-down biocontrol model of predator and parasitoid with insecticide treatment. This model shows that the combined effect of the enemies can stabilize the system. This is qualitatively seen to match the 2007-2013 time period data (figure 1 in (\cite{bahlai2015shifts}), with the emergence of the parasitic wasp (\emph{Aphelinus certus}) in 2006-2007.

All of these simulations are compared to ET and EIL. The combined biocontrol models are seen to be able to suppress pest populations to below these levels, whereas single enemy models, or only insecticide use, are unable to. This suggests a combination of enemies is more effective in pest control and should be a strategy in integrated pest management programs.
We also simulate the progression of events in the 2000-2013 period, where there is only a predator for aphid control in 2000-2004, but insecticide use gradually increases from 2005-2008, with parasitoids coming into the system in 2006-2007. The effect of the various models during the various time windows is seen to capture the actual data qualitatively (\cite{bahlai2015shifts}). 

Overall, these simulations demonstrate that biological control has the potential to significantly suppress soybean aphid populations. Adopting strategies that support beneficial insect populations may therefore reduce the need for insecticide-based management. Indeed, our simulation results indicate that having a mixed natural enemy population results in the highest levels of aphid suppression. Although further experimental and theoretical work may be necessary to validate these results, our findings do suggest that strategies that focus on preserving diverse natural enemy populations may be more efficacious than the augmentative release of a single natural enemy species. For example, increasing landscape complexity, by maintaining floral strips, a higher proportion of semi-natural habitats, and/or less-intensively managed farmland significantly increases natural enemy diversity and abundance (\cite{begg2017functional,jachowicz2025highly}). Decreasing agrochemical inputs, including early season systemic insecticide applications as well as foliar insecticide sprays, can also significantly boost natural enemy populations. However, this strategy for natural enemy conservation may be more difficult for growers to adopt; rescue insecticide spray treatments are sometimes necessary to address emergent pest issues and preserve yield. 

Also note, our bio control models are modeled via smooth functional responses, but non-smooth responses can be effective for control (\cite{banerjee2025two, parshad2021some}). Furthermore, while there is no Allee effect in the summer months for the pest, due to its life history strategy being parthenogenetic reproduction over the summer, there could be one in effect during winter, since reproduction is sexual on the winter host. Thus explorations of a ``time dependent" or ``time periodic" Allee effect could also be considered. Also, cannibalism, the killing and consumption of conspecifics, is commonplace in many ecosystems (\cite{polis1981evolution}). This has also been considered in some bio-control models for aphids(\cite{houdkova2006scaling}). Although Lacewing are cannibalistic, we do not focus on this life history strategy in depth, in the current manuscript. However, we do provide a preliminary model (see supplement). Further experimental work is needed to quantify the functional response accurately, in the presence of cannibalism. We plan to explore these aspects empirically in future work. The other aspect of cannibalism that can be considered from a modeling viewpoint is its effect on both predator and pest species (\cite{al2018exploring}).

As we continue to advance our predictive capability and ability to simulate aphid natural enemy dynamics, there are several important additional factors that need to be considered when developing these sustainable integrated pest management recommendations. In particular, it will be critical to consider aphid resistance dynamics. Previous work has shown that populations of soybean aphids have adapted to be resistant to both insecticides (\cite{hanson2017evidence}) as well as aphid-resistant \textit{Rag} soybean varieties (\cite{hill2012resistance}). While coexistence of these two biotypes on resistant soybean plants has been explored (\cite{banerjee2022exploring}), the effect of bio-control on these biotypes has not been investigated. It is possible that biological control may differentially affect resistant and susceptible aphids. For example, predators that forage on aphids on resistant soybean plants may experience diminished fitness (\cite{lundgren2009direct}). Previous simulations also suggest that parasitoids may alter virulent and avirulent population dynamics (\cite{hopper2023modeling}). Future modeling efforts should therefore consider differential responses to both insecticides and natural enemies between different aphid biotypes. Furthermore, with climate change and increased precipitation predicted to occur in the Midwest over the next few years, bio-control under such abiotic stresses should also be further investigated. For example, it has been recently shown that virulent biotypes do better than avirulent ones on flooded soybean plants (\cite{lewis2025host}). Again, investigating the effect of combined bio-control programs in such situations would be useful.

Additionally, our simulations suggest that insecticides do have the potential to alter aphid natural enemy population dynamics. In this study, we primarily considered the influence from mid-season foliar insecticide sprays. However, future work should also consider the impacts of early-season neonicotinoid seed treatments in detail. In soybeans, these seed treatments typically protect seedlings from aphids and other herbivores for the first 28 - 35 days after planting. Although neonicotinoid seed treatments are widely used in soybean production (\cite{douglas2015large}), multiple studies have found that they do not provide consistent yield benefits in soybean (\cite{mourtzinis2019neonicotinoid}). In addition to providing minimal protection from insect pests, it is also possible that these seed treatments can negatively impact beneficial insects. Aphids that develop on treated plants can potentially expose predators and/or parasitoids to neonicotinoid insecticides, in turn negatively impacting predator fitness (\cite{esquivel2020thiamethoxam}) and limiting potential biocontrol service. Experimentally quantifying these types of factors poses significant logistical challenges. Through future exploration of these interactions in modeling simulations, we will be able to identify the scenarios and agricultural practices that best support biological control. Although the work presented in this manuscript primarily focuses on soybeans and soybean aphids, it is likely that our models can also be applied more broadly to other systems.

\section*{Acknowledgments}

This work is supported by the Agricultural and Food Research Initiative grant no. 2023-67013-39157 from the USDA National Institute of Food and Agriculture. Dylan Murray and Cy Marchese, who assisted with rearing insects and data collection for laboratory experiments. Gary Snyder and Paul Ritter provided logistical support and assistance in maintaining experimental growth chambers for plant rearing and for performing experiments. 

\printbibliography


\section{Supplemental Materials}
\subsection{Supplemental Methods}

\subsubsection{\textbf{Insect Rearing for Laboratory Experiments}}
\label{supp_materials_1}
All assays were conducted using an avirulent, biotype 1 soybean aphid colony. The colony originated from the Soybean Aphid Biotype Stock Center at the University of Illinois (maintained by Doris Lagos-Kutz) and had been maintained under ambient laboratory conditions for approximately three years prior to starting the study. The colony was held at room temperature and under supplemental lighting (16L:8D).

\textit{Aphidius colemani} parasiotid wasps were originally obtained from NaturesGoodGuys Beneficial Insectary and used to start a colony; wasps were continuously reared on avirulent soybean aphids for 1 year under the same rearing conditions previously described. To increase humidity in the cages, we also added an approximately 15 x 30 cm tub of water to the bottom of each cage. The opening of the tub was covered with fine mesh fabric to prevent parasitoids from entering the water.

Green lacewing (\textit{Chrysoperla rufilabris}) eggs were obtained from Arbico Organics and reared to the third instar under laboratory conditions. We received the eggs glued onto a paper card, and monitored the card daily for larval emergence. Within 24 hours of larval emergence, larvae were individually transferred into 60 mm petri dishes that contained a moistened filter paper circle and one soybean leaf with avirulent soybean aphids. Petri dishes were provisioned with aphids every one to two days, ensuring that larvae had a continuous supply of prey. For all studies, we maintained the lacewing larvae on petri dishes with prey for 7-10 days, at which point we transferred larvae into fresh petri dishes with only a soybean leaf. Larvae were then starved for 24 hours prior to starting experiments.

\subsubsection{\textbf{Impact of Parasitism on Soybean Aphid Fecundity}}
\label{supp_materials}
\begin{figure}[H]
\begin{subfigure}{.42\textwidth}
\centering
\includegraphics[width=6cm, height=5cm]{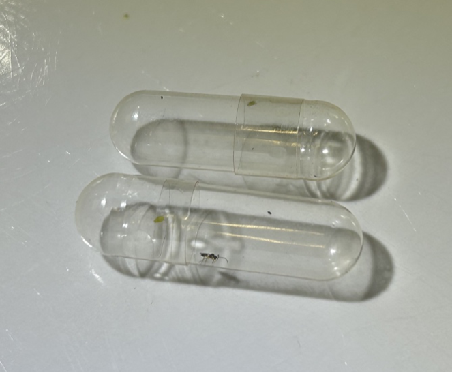}
\subcaption{}
\label{fig:supp_1}
  \end{subfigure}
  \begin{subfigure}{.42\textwidth}
  \centering
  \includegraphics[width= 6cm, height=5cm]{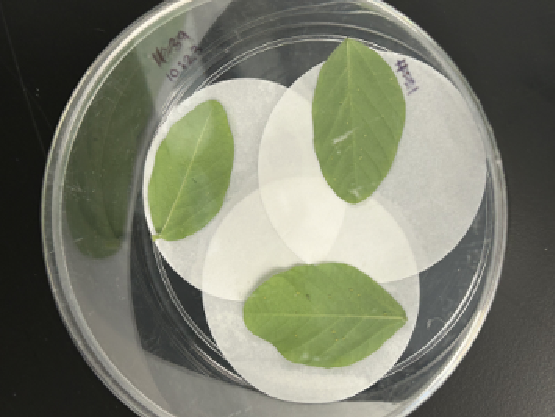}
  \subcaption{}
\label{fig:supp_2}
 \end{subfigure}
 \caption{Figure \ref{fig:supp_1}: Experimental set-up in aphid fecundity assays. Aphids were individually placed in $0.95$ mL gel capsules and exposed to either a female \textit{Aphidius colemani} (bottom) or no parasitoid as a control (top), and Figure \ref{fig:supp_2}: Experimental assay arenas used to measure parasitoid instar preference in section 
\eqref{Evaluating Parasitoid Preference for Different Aphid Instars}. 
Each leaf contained $10$ aphids belonging to one of three juvenile age groups, and a single female parasitoid wasp was placed into the center of each assay arena.}
\label{plot_supplemental_para}
 \end{figure}

To quantify parasitism impacts on aphid fecundity, we removed adult avirulent (Biotype $1$) soybean aphid adults (mixed ages) from a laboratory colony and placed them into individual $0.95$ mL gel capsules. Aphids were randomly assigned to one of two treatments. “Parasitized” aphids had one adult female \textit{Aphidius colemani} from a laboratory colony placed into the capsule, while  “Control” aphids had no parasitoid wasp added (Supplemental Figure \ref{fig:supp_1}). Parasitoid wasps were not age-standardized or subjected to controlled matings prior to use in the experiments. Instead, we monitored pairs of “Control” and “Parasitized” aphids simultaneously under a dissecting light microscope for a successful parasitoid attack (\textit{i.e.} the female wasp successfully probed the aphid body with her ovipositor). Replicates in which the parasitoid did not attack the aphid within five minutes were excluded from the experiment. Once an attack occurred, aphids from both treatments were then transferred to petri dishes for subsequent survival and fecundity monitoring. One observation (Parasitized – Rep $6$) was excluded from the final data analysis and summarization because the parasitoid attack was not successful - we monitored the aphid for $11$ days, and it never died or produced a parasitoid mummy.
\subsubsection{Impact of Cannibalism from Conspecific Predators}
\label{Impact of Cannibalism from Conspecific Predators}
We conducted functional response assays to evaluate how lacewing predation upon conspecifics might impact their consumption rates. These assays were conducted using petri dishes. Although confining predators and prey to a small petri dish rather than a whole plant can impact aspects of a predator’s functional response (\cite{messina1998host}), using petri dish assays allowed us to conduct larger numbers of experimental replicates simultaneously by removing limitations on plant and space requirements. Instead of modeling predator-prey interactions across a whole plant, these models represent the interactions that might occur within a smaller patch (i.e., a single aphid-infested leaf within a soybean plant). 

\vspace{0.2cm}

\par Each assay arena consisted of one $60$ mm petri dish; the bottom of the petri dish was covered with moistened filter paper to provide humidity throughout the course of the experiment. $35$ mm soybean leaf disks were placed in the center of each assay arena and infested with $3^\text{rd}$ – $4^\text{th}$ instar soybean aphid nymphs at one of seven densities: $2$, $4$, $8$, $16$, $32$, $64$, or $96$ aphids per leaf disk. Aphids were allowed to settle on the leaf disk for approximately $1$ hour prior to starting experiments.  

\vspace{0.2cm}

\par Two rounds of experiments were performed. In the first set, a single third-instar lacewing larva ($\sim 7$ days old) was added to each petri dish. Parafilmed petri dishes with aphids and lacewings were left for 24 hours, at which point we counted the number of aphids remaining. In the second set of experiments, three third-instar lacewing larvae were added to each petri dish, and we counted the number of aphids and lacewing larvae remaining after $24$ hours. To confirm that baseline predation rates were similar between the two experiments, we also included a single lacewing control, which consisted of petri dishes with one lacewing larva and $64$ aphids.

\vspace{0.2cm}

\par In the three-lacewing assays, we calculated the mean number of aphids that were eaten over a $24$-hour period (aphid consumption) and the number of lacewings remaining (cannibalism rate) at each initial aphid density. Linear models were fit using the lm() function in the R base package to evaluate how aphid consumption and cannibalism rates changed across varying initial aphid densities. Because the three-lacewing larvae and one-lacewing larvae assays were performed across separate periods of time, we did not make direct comparisons between the two sets of experiments.

\subsubsection{\textbf{Results for section \ref{Impact of Cannibalism from Conspecific Predators}}}
Data from the single lacewing larvae assays supported a Type II functional response curve $(\text{z} = -17.07, \ \text{P} < 0.001) $. Fitting the functional response using a Rogers Type II curve, the attack rate was estimated to be $3.54 \pm 0.320$ $(\text{z} = 10.75, \ \text{P} < 0.001)$, and the handling time was estimated to be $0.027 \pm 0.0013$ $(\text{z} = 20.36, \ \text{P} < 0.001)$. 

\vspace{0.2cm}

Contrasting the single lacewing larvae assays, we found no evidence for a Type II functional response when three lacewings were present. This may reflect the higher levels of consumption that occurred with the increase in predators, and it is likely that the range of initial aphid densities used in the assay was insufficient to capture a type II or Type III functional response. Overall, there was a positive linear relationship between aphid consumption and the initial aphid density ($\text{F}_{1,27} = 50.3, \text{P} < 0.001$, Table \ref{table:predator_experiment_up} and Supplemental Figure \ref{fig:supp_4}). Similarly, lacewing cannibalism rates declined as initial aphid densities increased ($\text{F}_{1,27} = 17.5, \text{P} < 0.001$), although the linear relationship was weaker (Supplemental Figure \ref{fig:supp_5}).

\subsubsection{Predator bio-control with cannibalism and stage structure}

\begin{equation}
\begin{split}
\label{eq:04}
\frac{dh}{dt} &= a x, \ h(0) =0 \\[0.2em] 
\frac{dx}{dt} &=  (r-h)x -\textcolor{black}{\dfrac{ a_x x y_{j}}{1+(a_xt_x)x+(a_jt_j)y_j}}, \ x(0) =x_{0}\\[0.2em]
\frac{dy_{j}}{dt} &= \textcolor{black}{\dfrac{\epsilon_x a_xxy_{j}} {1+(a_xt_x)x+(a_jt_j)y_j}} + \textcolor{black}{\dfrac{(\epsilon_j -1) a_jy_{j}^2} {1+(a_xt_x)x+(a_jt_j)y_j}} +\xi y_a -{\beta_j y_j}, \ y_{j}(0) =y_{j_{0}}\\[0.2em]
\frac{dy_{a}}{dt} &=  \textcolor{black}{\beta_j y_j} - \delta_a y_{a}, \ y_{a}(0) = y_{a_{0}}\\
\end{split}
\end{equation}
For model \eqref{eq:04},
the aphid population is again influenced only by predation from the juvenile stage of the predator, and there is no interaction with adult predators. However, the cannibalistic nature of the predator is incorporated here. Now, the functional response is modified in the sense that it not only depends on the prey but also depends on juvenile predator density $y_j$ (\cite{rudolf2007consequences}). Since the cannibalistic nature is only seen in the juvenile stage of the predator under consideration (lacewing larvae), the net effect - gain or loss because of cannibalism is reflected in the juvenile predator population equation. Following (\cite{rudolf2007consequences}), we also assume that $0\leq e_j \leq 1$ to avoid a situation where the predator could grow indefinitely without any prey present. As in model \eqref{eq:03},
juvenile predators mature into adults, contributing to the growth of the adult predator population $y_a$. A general description of the model parameters is given in Table \ref{table_pred_cann-params}. 
If $a_j = 0$ means there is no attack on the conspecifics, thus no cannibalism, and model \eqref{eq:04} reduces to model
\eqref{eq:03}.
\begin{singlespace}
\begin{table}[H]
    \centering
    \begin{tabular}{|l l|}
        \hline
        \textbf{Parameters} & \textbf{Definition} \\
        \hline
        \multicolumn{2}{|l|}{} \\
        $a$    &  scaling parameter relating aphid cumulative density to its dynamics\\
        $r$    & maximum potential growth rate of aphids\\
        $\xi$  & birth rate of new juveniles from adult predators\\
        $\beta_j$ & rate at which juvenile predator matures to the adult stage\\
         $\epsilon_x, \epsilon_j$ & conversion efficiency for aphids and conspecific prey, respectively\\
        $a_x, a_j$ & attack rate of juvenile predators on aphids and conspecifics, respectively\\
        $t_x, t_j$ & handling time spent for prey and conspecifics, respectively\\
        $\delta_a$  & natural mortality of adult predators \\
        \hline
    \end{tabular}
        \caption{Model parameter definitions for \eqref{eq:04}}
    \label{table_pred_cann-params}
\end{table}
\end{singlespace}
\subsection{Supplemental Figures and Tables}

\begin{figure}[H]
    \centering
\includegraphics[width=6cm, height=5cm]{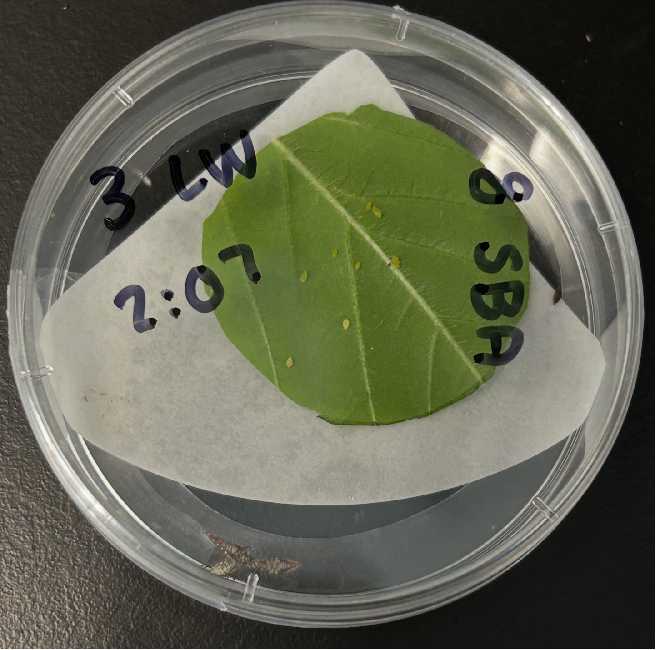}
\caption{Assay arena used for functional response assays. Three lacewing larvae can be seen cannibalizing each other at the bottom of the petri dish.}
\label{fig:supp_3}
\end{figure}

\begin{figure}[H]
    \centering
\includegraphics[width=7cm, height=5cm]{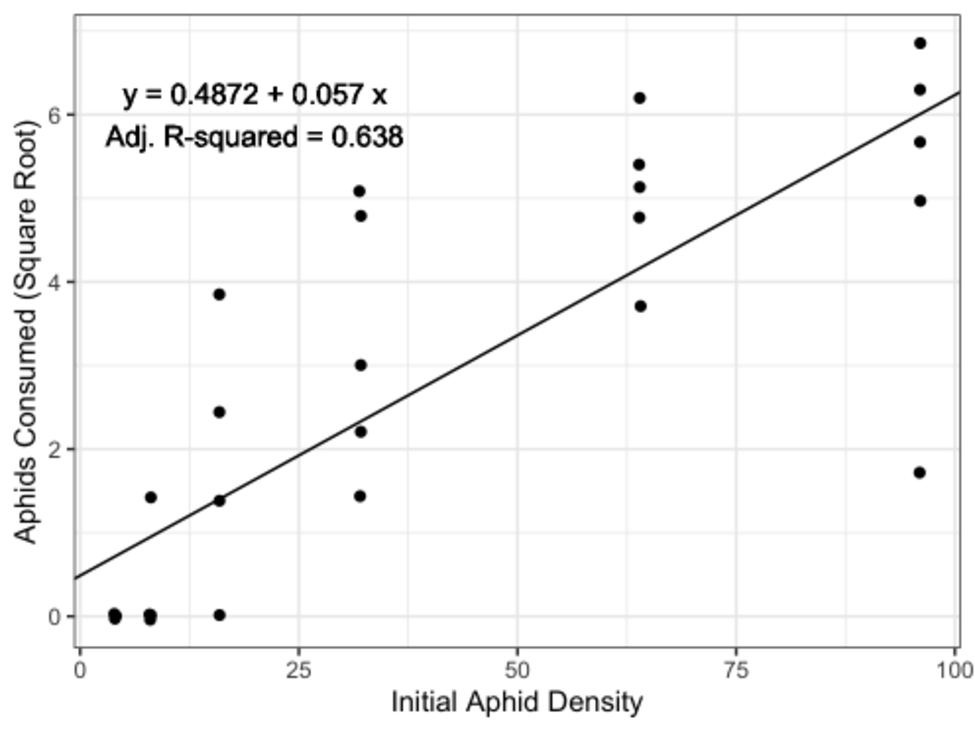}
\caption{Linear relationship between aphid consumption rate and initial aphid density in the three lacewing larvae assays. Data were square root transformed to address the assumptions of normality and heterogeneity of variance in the model residuals, and data are graphed using the square root transformation.}
\label{fig:supp_4}
\end{figure}

\begin{figure}[H]
\centering
\includegraphics[width=7cm, height=5cm]{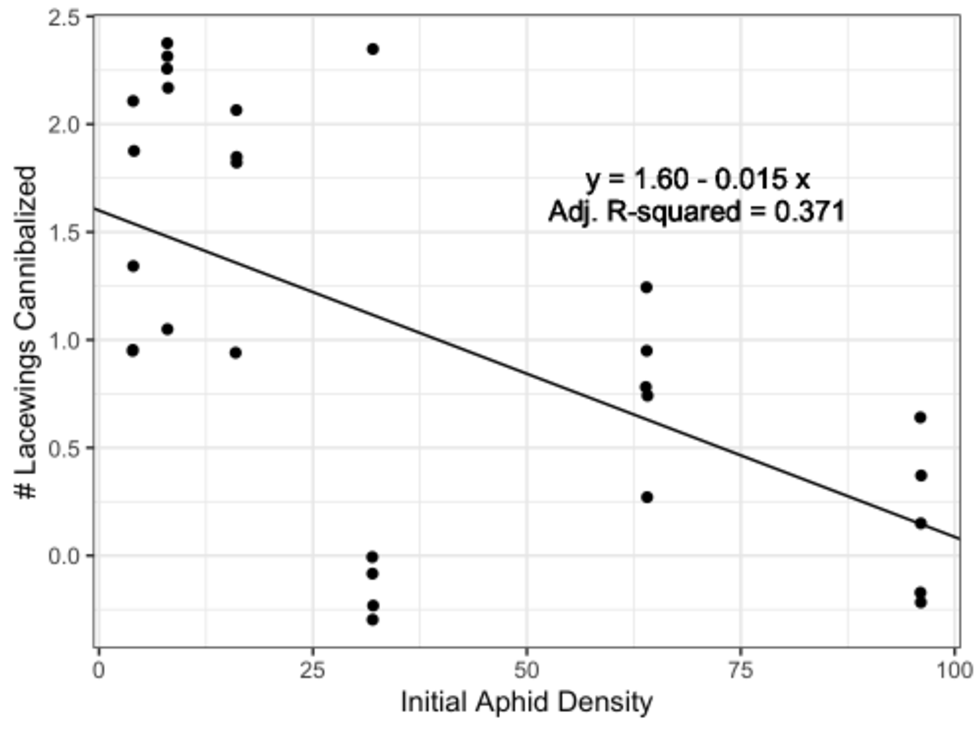}
\caption{Linear relationship between the number of lacewings consumed and initial aphid density in the three lacewing larvae assays.}
\label{fig:supp_5}
\end{figure}
\begin{singlespace}
\begin{table}[H]
\centering
\resizebox{\textwidth}{!}{
\begin{tabular}{|l|c|c|c|}
 \hline
\textbf{Initial Aphid Density} & \textbf{ \# Aphids Consumed (Mean $\pm$ SE)} & \textbf{\# Lacewings Consumed (Mean $\pm$ SE)} & \textbf{N} \\
\hline
4& 4.00 $\pm$ 0.00 & 1.40 $\pm$ 0.25 & 5\\
\hline
8 & 7.60 $\pm$ 0.40 & 1.80  $\pm$ 0.20 & 5\\
\hline
16& 11.0 $\pm$ 2.68 & 1.75 $\pm$ 0.25 & 5\\
\hline
32 & 19.6 $\pm$ 5.10 & 0.40  $\pm$ 0.40 & 5\\
\hline
64& 38.0 $\pm$ 3.90 & 0.80 $\pm$ 0.20 & 5\\
\hline
96 & 69.4 $\pm$ 7.29 & 0.20  $\pm$ 2.50 & 5\\
\hline
\end{tabular}}
\caption{Variation in aphid consumption rate (\# aphids consumed by lacewings over a $24$-hour period) and lacewing cannibalism rates (\# of lacewings that were consumed by conspecifics over a $24$-hour period) in the three-lacewing larvae assays. N denotes the number of experimental replicates.}
\label{table:predator_experiment_up}
\end{table}
\end{singlespace}

\begin{singlespace}
\begin{table}[H]
\centering
\begin{tabular}{|l|c|c|c|c|c|c|}
\hline
\textbf{Stage ($\mathbf{j}$)} & \textbf{I1} & \textbf{I2} & \textbf{I3} & \textbf{I4} & \textbf{Female} & \textbf{Male} \\
\hline
\textbf{Probability} & 1 & 0.9474 & 0.8889 & 1 & 0 & -- \\
\hline
$\mathbf{D(j)}$ & 0 & 1 & 2 & 0 & 16 & 0 \\
\hline
$\mathbf{N(j)}$ & 19 & 18 & 16 & 16 & 0 & 0 \\
\hline
$\mathbf{I(j)}$ & 19 & 19 & 18 & 16 & 16 & -- \\
\hline
\end{tabular}
\caption{Soybean aphid stage-specific survival rate and number from stage $j$ to $j+1$. For the last preadult stage (i.e., I4), it is the survival rate from the last preadult stage to adult (female and male); all aphids in these experiments are female, because we were using a parthenogenetic population. All aphids will die in the adult stage. Life table analyses were performed to determine stage-specific survival rates in soybean aphids. Briefly, neonate soybean aphid nymphs were individually caged on soybean plants and checked daily for (a) survival, (b) instar, and (c) daily nymph production once the aphid reached the adult life stage. Using the software TwoSex MS Chart, we were then able to calculate age-stage-specific survival rates. Any aphids that escaped from their cage were excluded from the analysis, and in total, we performed 19 successful experimental replicates.}
\label{transfer rates}
\end{table}
\end{singlespace}
\begin{flushleft}
\textbf{Notes:} \\
$D(j)$: the number of deaths occurred in stage $j$. \\
$N(j)$: the number of individuals in stage $j$ will survive to stage $j+1$. \\
$I(j)$: the number of individuals that entered stage $j$. \\
The probability is calculated as $N(j)/I(j)$. \\
The probability from pupa to female is 1 (i.e., $16/16$). \\
The probability from pupa to male is 0 (i.e., $0/16$).
\end{flushleft}

\begin{figure}[H]
\begin{subfigure}{.48\textwidth}
\centering
\includegraphics[width=7cm, height=5cm]{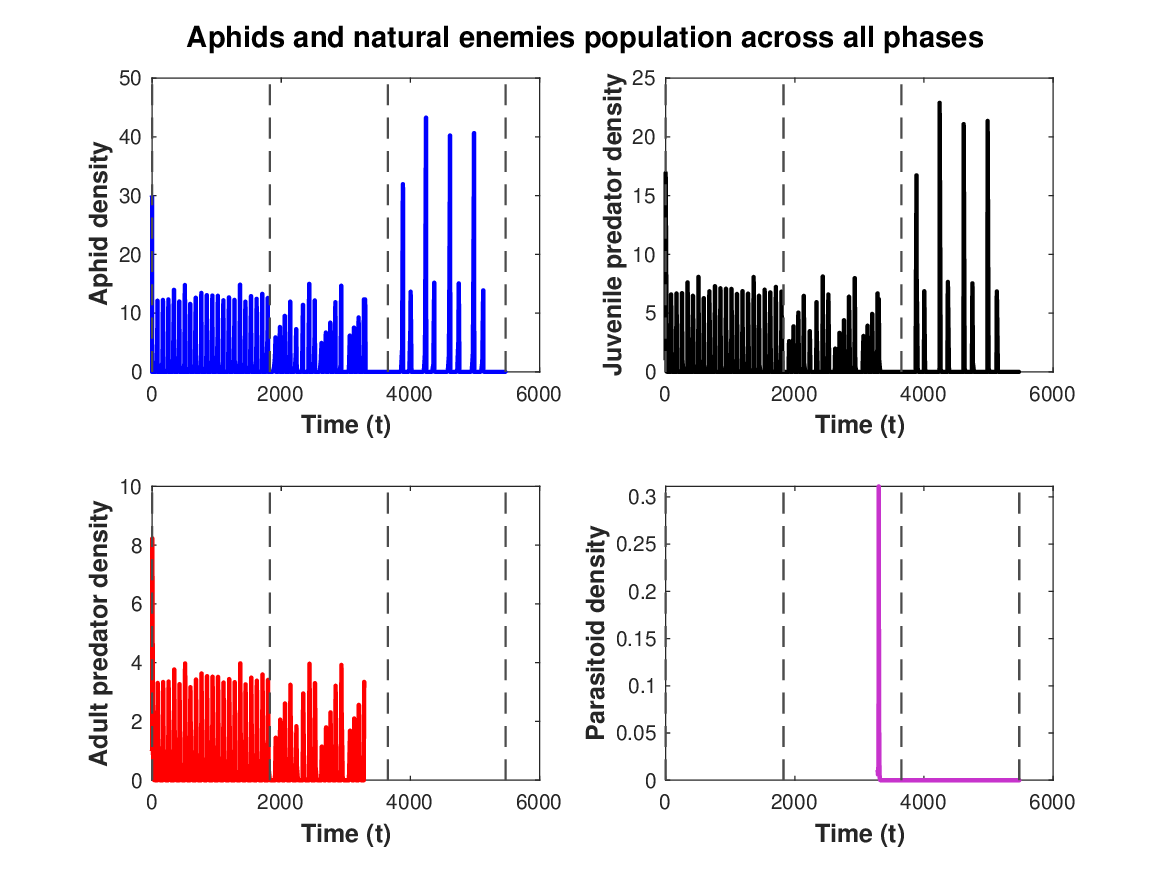}
\subcaption{Population dynamics across phases}
\label{fig:combined_ts}
  \end{subfigure}
  \begin{subfigure}{.48\textwidth}
  \centering
  \includegraphics[width= 7cm, height=5cm]{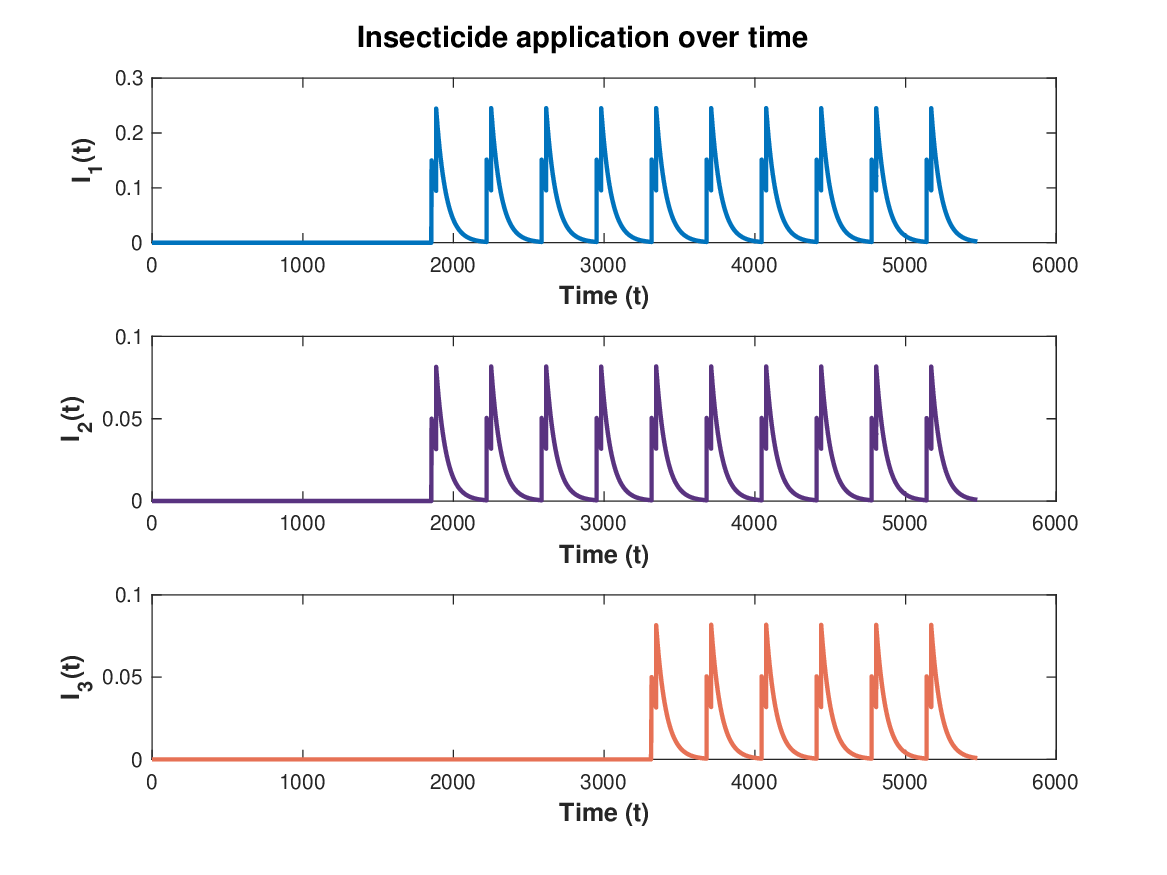}
  \subcaption{Re-application of insecticide}
\label{fig:combined_insec}
 \end{subfigure}
 \caption{This time series corresponds to Figure 
 \ref{combined_populations}.
 The time series plot is divided into three different phases, representing the population dynamics in each phase. The initial condition in the first phase was $[0,30,1,1]$, and then the initial condition for the next models is carefully chosen from the end value of the previous phase. The parameters used are $ a=0.000005, r=0.3, \alpha=0.012,\beta_1=0.6,\beta_2=\beta_3=0.45,\gamma_1=0.2,\gamma_2=0.38,\eta_1=0.5875,\eta_2=0.5667,\epsilon_1=\epsilon_2=\epsilon_x=0.55,q_1=q_2=0.6,a_x=a_1=a_2=1.6281,t_x=t_1=t_2=0.0260,\beta_j=0.4,\delta_a=\delta_p=\delta_j=0.3,c=\text{log}(2)/45,\delta_1=0.15,\delta_2=\delta_3=0.05$.}
 \label{combined_ts_insec}
 \end{figure}

\begin{figure}[H]
 \begin{subfigure}{.48\textwidth}
\centering
 \includegraphics[width=7cm, height=5cm]{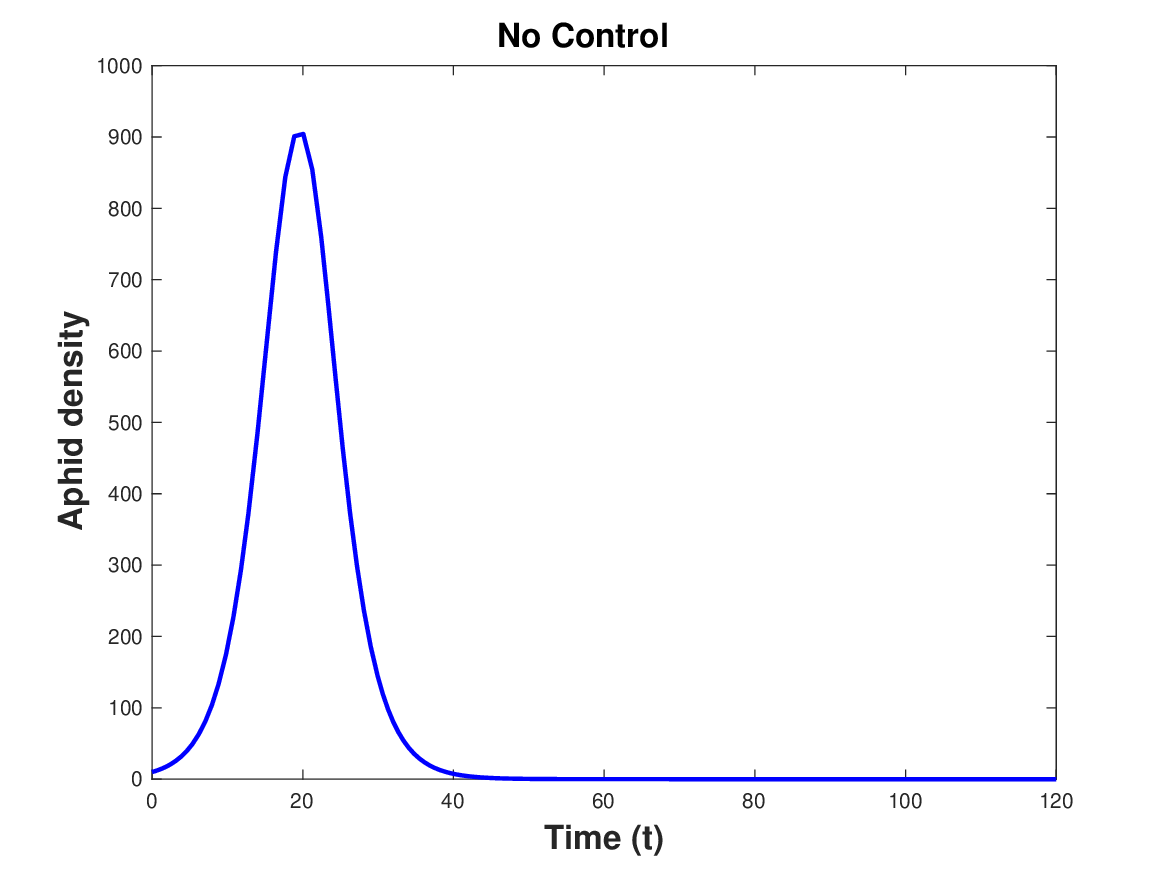}
 \subcaption{ }\label{fig:supp_first_no_average_para}
  \end{subfigure}
  \begin{subfigure}{.48\textwidth}
  \centering
  \includegraphics[width= 7cm, height=5cm]{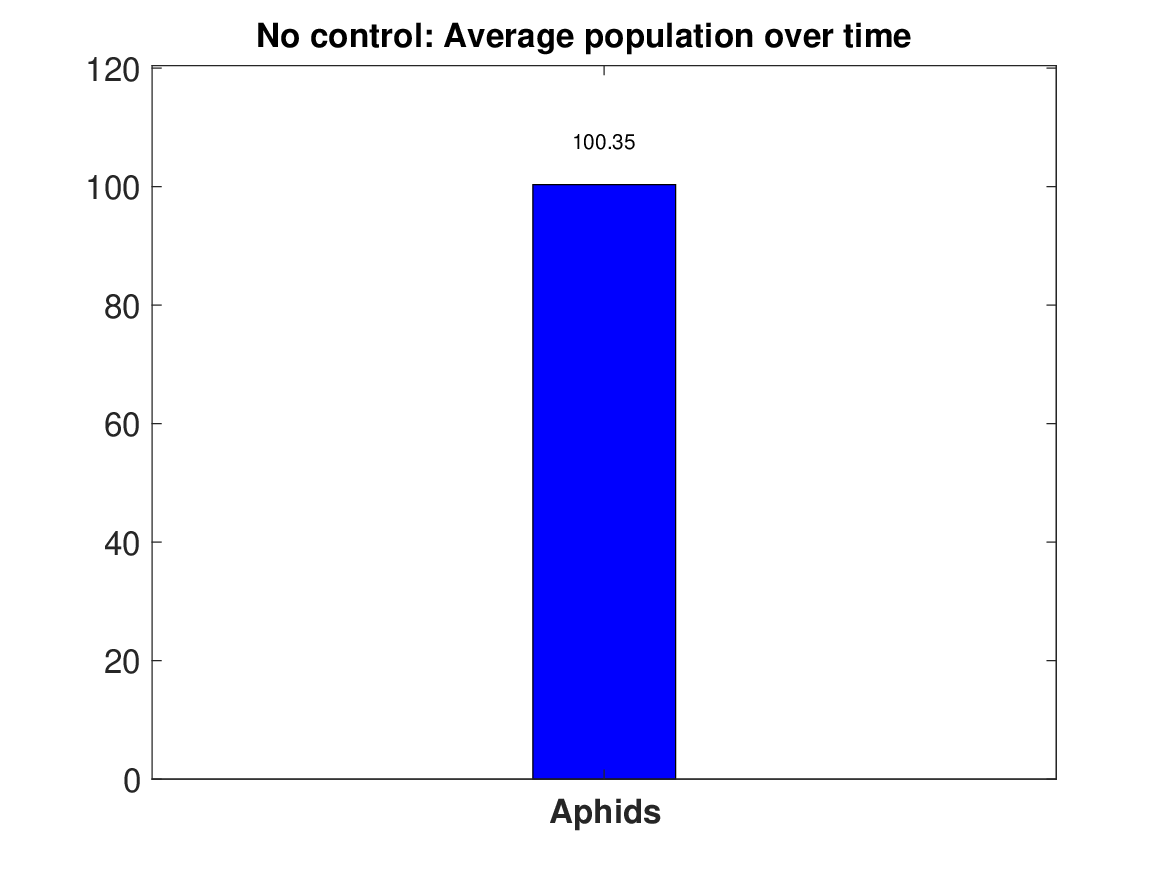}
  \subcaption{}
\label{fig:supp_second_no_average_para}
 \end{subfigure}
 \caption{These subplots show \ref{fig:supp_first_no_average_para} changes in aphid population density over time and \ref{fig:supp_second_no_average_para} the average aphid population for model 
 \eqref{eq:0}, 
 which corresponds to a single-species aphid model with no control. The scaling parameter and aphid growth rate are fixed at $a= 5 \cross 10^{-5}, r=0.3$ with initial density as $[0,10]$.}
\label{supp_plot_average_no_control}
\end{figure}
 
\begin{figure}[H]
  \begin{subfigure}{.48\textwidth}
\centering
  \includegraphics[width=7cm, height=5cm]{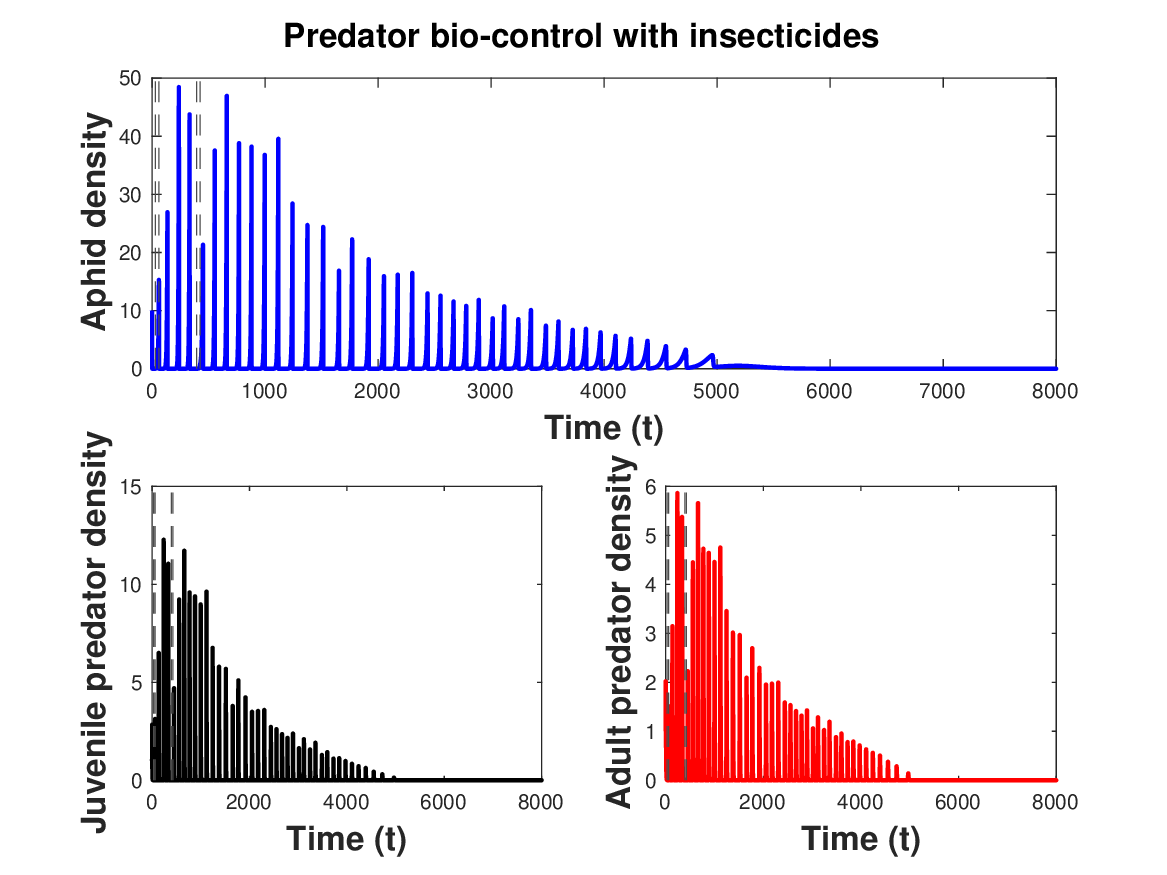}
 \subcaption{ }
 \label{supp_pred_ts_1}
  \end{subfigure}
  \begin{subfigure}{.48\textwidth}
  \centering
  \includegraphics[width= 7cm, height=5cm]{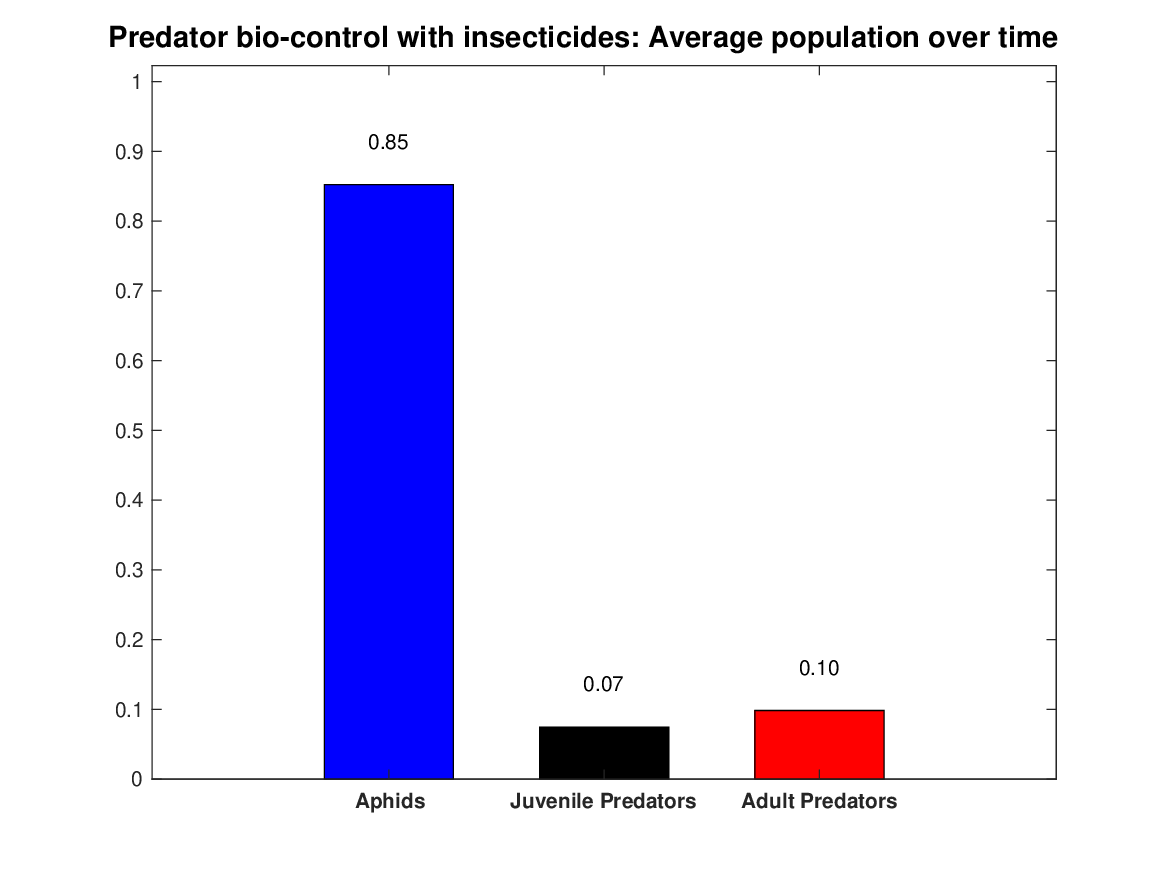}
  \subcaption{}
\label{supp_pred_ts_2}
 \end{subfigure}

 \caption{ These subplots show \ref{supp_pred_ts_1} changes in aphid and predator population density over time and \ref{supp_pred_ts_2} the average aphid and predator population over a single field season for model,
 \eqref{eq:insectide04}, 
 which represents the predator biocontrol with insecticide application. The blue bar chart represents the average aphid population, while the black and red bar chart shows the average populations of juvenile and adult predators, respectively. The insecticide was sprayed two times in a year till $2^\text{nd}$ year, giving the spray set as $t=[30, 60, 395, 425]$. It can be seen that the population shows transient dynamics for a long time before stabilizing. The parameters used are $ a= 5 \cross 10^{-5}, r=0.3, e_x=0.25, a_x=1.6281,t_x= 0.0260,\beta_j=0.4,\xi=0, \delta_a=0.3, c=\text{log}(2)/45, \delta_1=0.1,\delta_2=\delta_3=0.05$ with I.C.$=[0,10,1,1]$.
 }
\label{supp_plot_average_pred_control}
 \end{figure}

  \begin{figure}[H]
  \begin{subfigure}{.32\textwidth}
\centering
  \includegraphics[width=5.8cm, height=5cm]{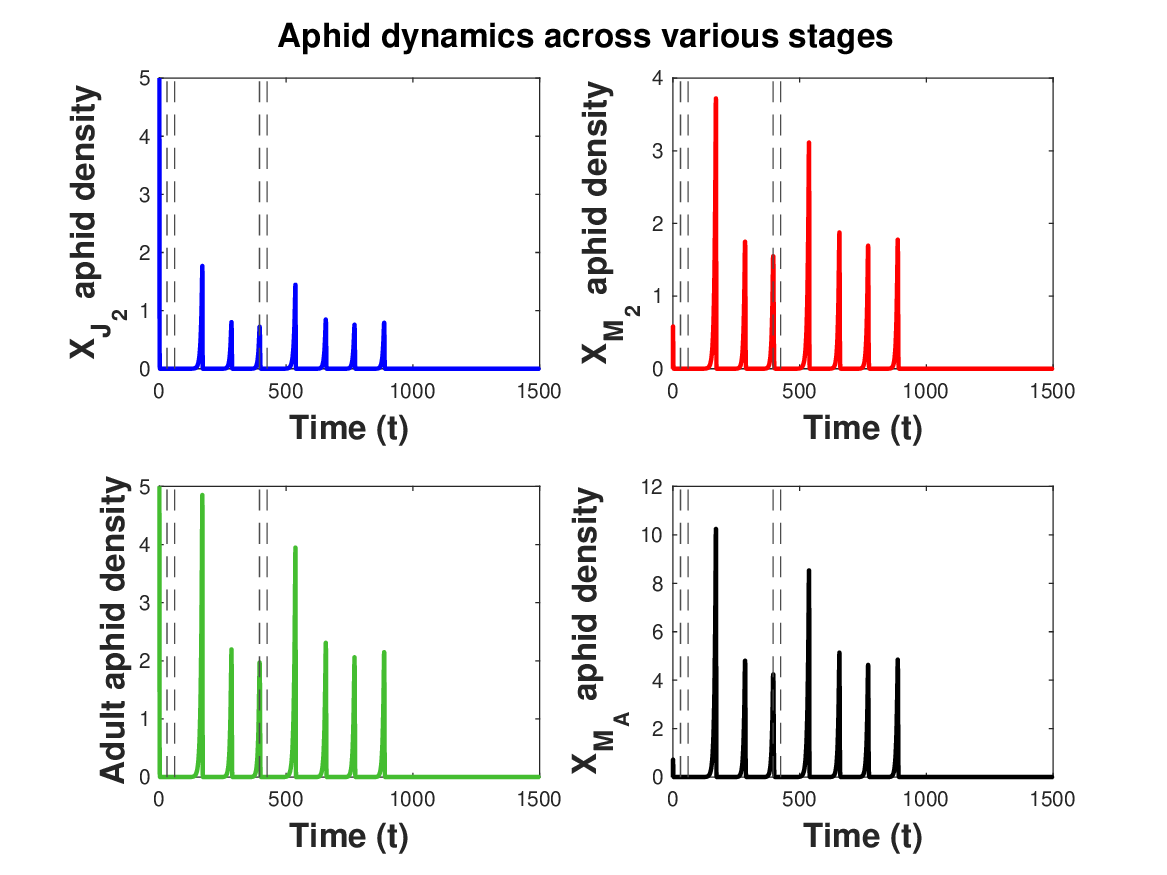}
 \subcaption{ }
\label{fig:supp_first_entry_average_para}
  \end{subfigure}
  \begin{subfigure}{.32\textwidth}
  \centering
  \includegraphics[width= 5.8cm, height=5cm]{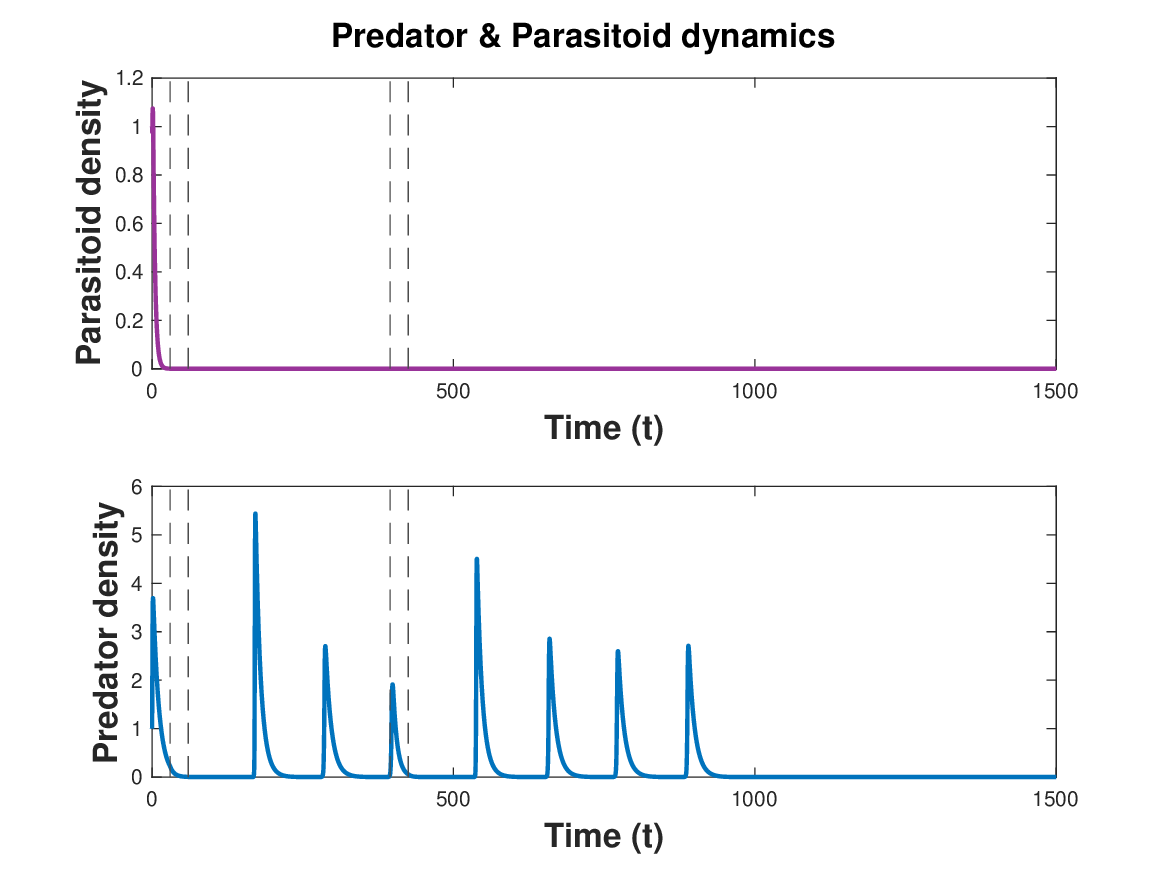}
  \subcaption{}
\label{fig:supp_second_entry_average_para}
 \end{subfigure}
  \begin{subfigure}{.32\textwidth}
  \centering
  \includegraphics[width= 5.8cm, height=5cm]{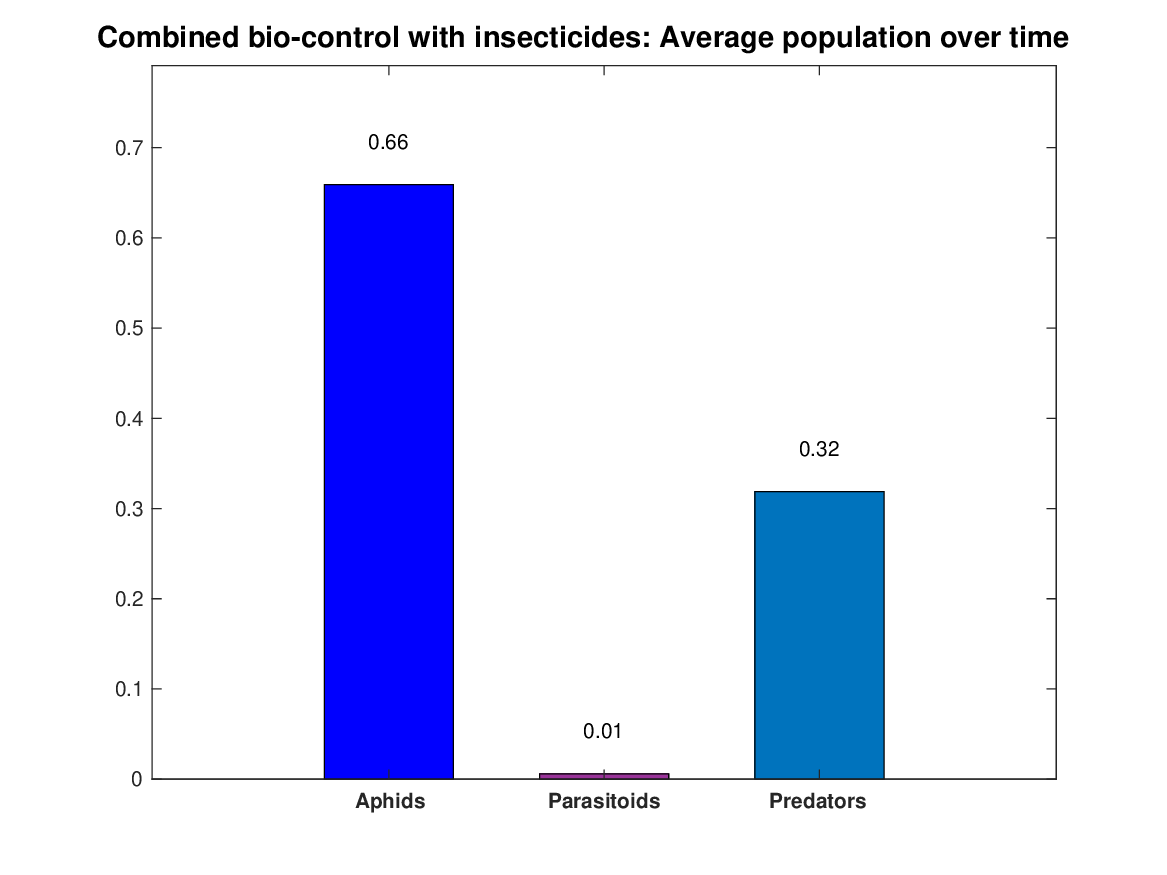}
  \subcaption{}
\label{fig:supp_third_entry_average_para}
 \end{subfigure}
 \caption{These subplots shows \ref{fig:supp_first_entry_average_para} changes in aphid population density across various stages over time, \ref{fig:supp_second_entry_average_para} changes in predator and parasitoid population density over time and, \ref{fig:supp_third_entry_average_para} shows the average aphid, parasitoid and predator population over a single field season 
 \eqref{eq:09}, 
 which accounts for the predator-parasitoid coexistence with insecticide treatment control strategy. The bar chart represents the average population over time as follows: blue for the aphids across all the juvenile and mummified stages, violet for the parasitoid population, and light blue for the predator population. The insecticide was sprayed two times in a season until $2^\text{nd}$ year, giving the spray set once at $t=30, 60, 395, 425$. It can be seen that the combined strategy stabilizes the aphid population across all stages when compared to Figure \ref{supp_plot_average_pred_control}. The parameters used are $ a= 5 \cross 10^{-5}, r=0.3, \alpha=0.012,\beta_1=0.6,\beta_2=\beta_3=0.45,\gamma_1=0.2,\gamma_2=0.38,\eta_1=0.5875,\eta_2=0.5667,\epsilon_1=\epsilon_2=0.25,q_1=q_2=0.6,a_1=a_2=1.6281,t_1=t_2=0.0260,\delta_p=0.1,\delta_j=0.3,\delta_1=0.1,\delta_2=\delta_3=0.05,c=\text{log}(2)/45$ with I.C.$=[0,5, 0, 5, 0, 1, 1]$.}


\label{supp_plot_average_para_control}
 \end{figure}
 
\end{document}